\documentclass[aps,showpacs,nobibnotes,floats,eqsecnum,nofootinbib]{revtex4}
\usepackage{amsmath,amsfonts,amssymb,amsthm}           
\usepackage{hyperref}                                  
\usepackage{graphicx}                                  

\newtheorem{assertion}{Assertion}                      
\newcommand{\bvec}[1]{\boldsymbol{#1}}                 
\newcommand{\pd}{\partial}                             
\newcommand{\Res}[2]{\mathtt{Res}\left[#1, #2\right]}  
\newcommand{\cket}[1]{\ensuremath{\left|{#1}\right>}}  
\newcommand{\brac}[1]{\ensuremath{\left<{#1}\right|}}  
\newcommand{\bracket}[2]{\left<{#1}|{#2}\right>}       
\newcommand{\bigO}{\mathcal{O}}                        
\newcommand{\smallO}{\mathit{o}}                       
\newcommand{\twoCol}[2]{\begin{pmatrix}{#1}\\{#2}\end{pmatrix}} 

\newcommand{\twoCases}[4]
{
  \left\{
  \begin{array}{ll}
    {#1}, & {#2}, \\
    {#3}, & {#4}
  \end{array}
  \right.
}

\def\hc{^\dagger}                               
\def\cc{^*}                                     
\def\Lagr{\mathcal{L}}                          
\newcommand{\Ints}{\mathbb{Z}}                  
\newcommand{\Complex}{\mathbb{C}}               
\newcommand{\Reals}{\mathbb{R}}                 

\DeclareMathOperator{\rot}{rot}                 
\DeclareMathOperator{\divg}{div}                
\DeclareMathOperator{\IIm}{\mathtt{Im}}         
\DeclareMathOperator{\RRe}{\mathtt{Re}}         
\DeclareMathOperator{\sgn}{\mathtt{sgn}}        

\begin{document}
\title{Casimir Effect within $D=3+1$ Maxwell-Chern-Simons Electrodynamics}
\author{O.~G.~Kharlanov}\email{okharl@mail.ru}
\author{V.~Ch.~Zhukovsky}\email{zhukovsk@phys.msu.ru}
\address{Department of Theoretical Physics, Moscow State University,
         119991 Moscow, Russia}%
\pacs{12.60.-i, 12.60.Cn, 11.10.Ef, 03.65.Pm}

\begin{abstract}
Within the framework of the (3+1)-dimensional Lorentz-violating extended electrodynamics including the CPT-odd
Chern-Simons term, we consider the electromagnetic field between the two parallel perfectly conducting plates.
We find the one-particle eigenstates of such a field, as well as the implicit expression for the photon energy
spectrum. We also show that the tachyon-induced vacuum instability vanishes when the separation between the
plates is sufficiently small though finite. In order to find the leading Chern-Simons correction to the vacuum
energy, we renormalize and evaluate the sum over all one-particle eigenstate energies using the two different
methods, the zeta function technique and the transformation of the discrete sum into a complex plane integral
via the residue theorem. The resulting correction to the Casimir force, which is attractive and quadratic in the
Chern-Simons term, disagrees with the one obtained in [M.~Frank and I.~Turan, Phys. Rev. D \textbf{74}, 033016
(2006)], using the misinterpreted equations of motion. Compared to the experimental data, our result places a
constraint on the absolute value of the Chern-Simons term.
\end{abstract}
\maketitle

\section{Introduction}\label{sec:Intro}
For today, the Standard Model is proved with a convincing variety of
experiments. However, this theory cannot shed light on some essential aspects
of our understanding of the Nature, which still remain obscure and are of
great interest. Among them is the \emph{quantum gravity}, which is believed to
play a major role at the \emph{Planck scale of energies} ($E_{\text{Pl}} \sim
10^{19}$GeV). The Standard Model fails to incorporate the General Relativity
at the quantum level, since the quantization methods adopted in the former
theory lead to a nonrenormalizable quantum gravity in this case. Such theory
is unacceptable as a fundamental theory. At the same time, there exist some
candidates for such Fundamental theory, string theories, for instance, taking
the form of the Standard Model and the General Relativity in appropriate
low-energy limits. Thus searching for signatures of the Planck-scale physics
at experimentally attainable energies would be a natural way to choose between
these candidate theories, to constrain their parameters, and to clarify the
essence of the quantum gravity.

Planck energies being far from experimental attainment, the \textit{Standard Model Extension (SME)} was
elaborated. It is an effective theory (applicable at the energies $E \ll E_{\text{Pl}}$) formulated
axiomatically as a set of corrections to the Lagrangian of the Standard Model, which fulfill some ``natural''
requirements \cite{ColladayVAK:SME, Bluhm:SME}, such as \emph{observer Lorentz invariance, 4-momentum
conservation, unitarity, and microcausality}. In what follows, we will focus on a subset of the SME referred to
as the \textit{minimal} SME in \textit{flat} Minkowsky spacetime, in which local $SU(3)_C\times SU(2)_I\times
U(1)_Y$ gauge invariance and power-counting renormalizability are also required. A spectacular feature of such
requirements is that they reduce the diversity of possible correction terms down to a \textit{finite} number of
them. Each of them consists of a complex (pseudo)tensor constant (SME coupling) contracted with conventional
Standard Model fields and their spacetime derivatives. These constants are believed to stand for vacuum
expectation values of the fields featuring in the hypothetic Lorentz-covariant Fundamental theory and condensed
at low energies due to the spontaneous symmetry breaking mechanism. Indeed, it has been shown that such Lorentz
symmetry breaking can occur in some theories beyond the Standard Model \cite{SpontLCPTBreaking:1,
SpontLCPTBreaking:2, SpontLCPTBreaking:3, SpontLCPTBreaking:4}, leading subsequently to the SME. The SME can
thus be used to reduce the complexity of these theories and related calculations in the low-energy limit. It
also provides a standard for representation of the data obtained in experiments searching for Lorentz violation.

Recently, a number of theoretical researches has been performed aimed at investigating the vacuum structure of
this model (see, e.g., \cite{VAKLehnert, EbZhRazum, Andrianov, Altschul:PV, Jackiw}), and to study the assumed
Lorentz violation on various high-energy processes \cite{VAKPickering, AdamKlink, Altschul:SR, Lena,
Andrianov:0907}. Such effects as the vacuum photon splitting, vacuum birefringence, and vacuum Cherenkov
radiation were studied. The effect of Lorentz violation on the synchrotron radiation \cite{AltschulSR, IgorSR}
and the atomic spectrum and radiation properties \cite{BluhmHydrogen, BluhmHydrogen2, OlegHydrogen} were also
analyzed. As a result of such theoretical investigations and subsequent precise experimental tests, the new
constraints on Lorentz violation were placed.

On the other hand, even in the conventional Standard Model, the quantum structure of the vacuum manifests itself
in such effects as the Casimir effect \cite{Casimir, Lifshitz}, that have been directly observed. This effect
has been studied thoroughly within the Standard Model, including different approaches \cite{MiltonLectures,
Saharyan} (various regularization schemes, the calculations via the dyadic Green function and the vacuum
energy), setups (two parallel plates, a sphere, a plate and a sphere; non-perfect conductor plates, finite
temperature, fermion Casimir effect, etc.). The so-called \emph{Maxwell-Chern-Simons (MCS) electrodynamics} in
$(2+1)$ dimensions was also considered in this concern \cite{Milton}. The Casimir effect was also studied within
the context of extra dimensions (see, e.g., \cite*{Linares, KirstenFulling, Turan:CasimirExtDim}) and curved
space \cite{Toms_Ads5, Saharyan_deSitter, Casimir_EinsteinUniv}. Some attention was paid to this problem within
the SME, namely, within Extended Quantum Electrodynamics (see Sec.~\ref{sec:MCSQED}) \cite{Turan}, which, in a
certain particular case, takes the form quite analogous to the MCS electrodynamics.

Although the original version of Casimir effect concerned electromagnetic
field, this effect is much more general. It can be defined as the stress
(force per unit area) on the bounding surfaces when a quantum field is
confined within a finite volume of space. The boundaries can be material media
and also the interfaces between different phases of the vacuum or topologies
of space.

In our investigation, we focus on the extended electrodynamics, in order to
find the signature of the Lorentz violation in the electromagnetic Casimir
effect.

Our paper is organized as follows. In Sec.~\ref{sec:MCSQED}, we analyze the pure-photon sector of a particular
case of the SME, namely, of the $(3+1)$-dimensional Maxwell-Chern-Simons electrodynamics, in order to obtain the
expression for the vacuum energy within this theory. We prove directly that the Casimir force in such a theory
remains gauge-invariant although the energy-momentum tensor does not. In Sec.~\ref{sec:PhotonEigenStates}, we
focus on finding the eigenstates and the energy spectrum of the photon between the two parallel perfectly
conducting plates within the MCS electrodynamics. The one-photon energy spectrum, which follows from the
implicit expression \eqref{spec_cond} found in Sec.~\ref{sec:PhotonEigenStates}, is then analyzed in
Sec.~\ref{sec:SpectrumNotes}, where we show that when the plates are close enough to each other, the
imaginary-energy (tachyonic) states are negligibly few, so the instability of the theory is also negligible (see
assertion \ref{FinalAssertion}). In Sec.~\ref{sec:ZetaReg}, using approximate energy eigenvalues, we find the
leading correction to the zeta function corresponding to the one-photon energy-squared operator, and, in turn,
the leading correction to the Casimir energy. The contribution of the ``quasi-zero'' modes, which become trivial
in the Maxwell case, to the Casimir force, is also discussed. In Sec.~\ref{sec:ResidueTheorem}, we use the
strict approach based on the residue theorem \cite{ComplexAnalysis} to explicitly sum the vacuum energy series,
which contains the zeros of the transcendental function \eqref{spec_cond}. We then renormalize the resulting
complex plane integral and find the exact convergent expression \eqref{E_vac_ren} for the real part of the
Casimir energy. Section \ref{sec:Conclusion} summarizes the results obtained, and we discuss the constraints
they place on the parameters of the Lorentz violation.

\section{Extended Maxwell-Chern-Simons electrodynamics}\label{sec:MCSQED}
\subsection{General notes}
After the electroweak symmetry is broken within the minimal SME, the extended Quantum Electrodynamics arises
\cite{ColladayVAK:SME}. For one fermion generation, it contains 10  additional (tensor) coupling constants which
introduce the interaction between the Dirac electron, the photons, and the condensates of Planck-scale fields
whose nonzero values violate Lorentz invariance at low energies:
\begin{eqnarray}
  \Lagr_{\text{ext.QED}} &=& -\frac14 F_{\mu\nu}F^{\mu\nu} + \bar\psi(i \Gamma^\nu D_\nu - M)\psi
                             -\frac14 (k_F)_{\mu\nu\alpha\beta}F^{\mu\nu}F^{\alpha\beta}
                             +\eta^\mu A^\nu \tilde{F}_{\mu\nu},\\
  \Gamma^\nu &=& \gamma^\nu + c^{\mu\nu} \gamma_\mu + d^{\mu\nu} \gamma_5\gamma_\mu + e^\nu + i f^\nu\gamma_5 +
                             \frac12 g^{\alpha\beta\nu}\sigma_{\alpha\beta},\\
  M &=&          m  + a_\mu \gamma^\mu + b_\mu \gamma_5\gamma^\mu + \frac12 H_{\mu\nu}\sigma^{\mu\nu},
\end{eqnarray}
where $\tilde{F}_{\mu\nu} = \frac12 \epsilon_{\mu\nu\alpha\beta}F^{\alpha\beta}$ is the dual field strength
tensor, $\epsilon_{\mu\nu\alpha\beta}$ is the Levi-Civita symbol, with $\epsilon^{0123} = -\epsilon_{0123} =
+1$, $D_\nu = \pd_\nu + i e A_\nu$ is the $U(1)$-covariant derivative, and $m$ and $e$ are the electron mass and
charge, respectively. $\gamma^\mu$, $\gamma_5$, and $\sigma_{\mu\nu}$ denote the Dirac matrices, while spacetime
indices $\mu,\nu,\alpha,\beta = 0,1,2,3$. For the Lorentz-violating couplings to be unambiguously defined, they
should be real traceless tensors with certain symmetry properties, such as $H_{\nu\mu} = -H_{\mu\nu}$.

It should be stressed that the theory remains
\emph{observer-Lorentz-invariant}, since the coupling constants are tensors
under the boosts and rotations of the reference frame. This means that the
Lagrangian is a scalar, and physics in different reference frames is
equivalent though not the same, because the values of the components of the
coupling constants depend on the choice of the frame. Nevertheless, these
components in different frames also differ by the Lorentz transformation,
hence, if one takes the latter one into account, physics in different
reference frames becomes fully Lorentz-covariant.

In a fixed reference frame, however, the existence of the Lorentz violation
affects, e.g., how the energy value of a free particle depends on its
momentum, in such a way that its energy-momentum vector is no more a 4-vector
under rotations and boosts of the \emph{momentum} (\emph{active} Lorentz
transformations). In other words, extended electrodynamics, as well as the SME
itself, is a theory with broken \textit{particle} Lorentz invariance and
preserved \textit{observer} Lorentz invariance \cite{ColladayVAK:SME}.

Most of the Lorentz-violating couplings are tightly constrained (see, e.g.,
\cite{Bluhm:SME} and references therein); some of them can be excluded from
the theory using certain unitary transformations, e.g., $a_\mu$ vanishes if we
make a transformation $\psi \to e^{-ia_\mu x^\mu}\psi$ \cite{ColladayVAK:SME}.

In the present paper, we will study the contribution of the axial vector
coupling $\eta^\mu$ into the electromagnetic (i.e., boson) Casimir effect
between the two parallel conducting plates, i.e., within the theory with the
Lagrangian
\begin{eqnarray}
    \Lagr &=& -\frac14 F_{\mu\nu}F^{\mu\nu} + \bar\psi(i\gamma^\mu D_\mu - m) \psi + \Lagr_{\text{CS}}, \label{L_MCS4}\\
    \Lagr_{\text{CS}} &=& \frac12 \eta^\mu \epsilon_{\mu\nu\alpha\beta} A^\nu F^{\alpha\beta}.
\end{eqnarray}
The CPT-odd Lorentz-violating correction $\Lagr_{CS}$ entering the pure-photon
sector of the theory has the form of the so-called \emph{Chern-Simons (CS)
term} \cite{ChernSimons}. Within the context of electrodynamics, it is also
called after Carroll, Field, and Jackiw who have shown that this term can be
induced by axions \cite{CarrollFieldJackiw}. It can also be a manifestation of
a nonzero background torsion \cite{CSViaTorsion}.

The theory with Lagrangian \eqref{L_MCS4} can be treated as a four-dimensional
analogue of the so-called Maxwell-Chern-Simons (MCS) electrodynamics
\cite{ChernSimons}, which is usually considered in three dimensions and has
the following Lagrangian:
\begin{equation} \label{L_MCS3}
    \Lagr_{\text{(2+1)D MCS QED}} = -\frac14 F_{\mu\nu}F^{\mu\nu}
                                    +\bar\psi(i\gamma^\mu D_\mu - m) \psi
                                    + \frac{\eta}{2} \epsilon_{\mu\alpha\beta} A^\mu F^{\alpha\beta},
\end{equation}
where $\epsilon_{\mu\alpha\beta}$ is the three-dimensional Levi-Civita symbol, and $\mu,\nu,\alpha,\beta =
0,1,2$. Due to this analogy, we will refer to the former theory as to the \emph{four-dimensional
Maxwell-Chern-Simons electrodynamics}. In (2+1) dimensions, the CS term does not violate Lorentz invariance,
since $\eta$ is a pseudoscalar, and, when it is present, the free electromagnetic field $F_{\mu\nu}$ satisfies
the Klein-Gordon equation with a nonzero mass equal to $2|\eta|$ \cite{ChernSimons}. However, the gauge
invariance, which is typically absent for massive gauge fields, survives for nonzero $\eta$, since the
Chern-Simons term is changes by a total derivative under gauge transformations.

The Chern-Simons term within (3+1)-dimensional electrodynamics has slightly different properties. It does also
change by a total derivative under $U(1)$ gauge transformations, and the action remains gauge-invariant (if
appropriate boundary conditions are fulfilled):
\begin{eqnarray}
  A_\mu      &\to& A_\mu - \pd_\mu\alpha(x),  \label{gauge_A}\\
  \psi       &\to& e^{i e \alpha(x)} \psi,  \\
  \bar\psi   &\to& e^{-i e \alpha(x)}\bar\psi,\\
  \Lagr &\to& \Lagr - \pd_\nu\left(\frac{\eta_\mu}{2} \alpha(x)  \epsilon^{\mu\nu\alpha\beta}F_{\alpha\beta}\right).
\end{eqnarray}
However, first, the coupling constant $\eta^\mu$ is a \emph{pseudovector}, and
thus violates Lorentz invariance. In particular, it can cause a spatial
anisotropy of the theory. Second, if $\eta^\mu \ne 0$, the theory can become
unstable \cite{ColladayVAK:SME,CarrollFieldJackiw}. We will discuss this issue
in Sec.~\ref{sec:SpectrumNotes}, and now we will consider the 4-dimensional
MCS electrodynamics in more detail.

Varying the action $\mathcal{A} = \int\Lagr\, d^4x$ with respect to $A_\mu$,
$\psi$, and $\bar\psi$, we obtain the equations of motion
\begin{eqnarray}
  \pd_\mu F^{\mu\nu} + 2 \eta_\mu \tilde{F}^{\mu\nu} &=& j^\nu,  \label{MCS_eqmotion_A}\\
  (\gamma^\mu(i\pd_\mu - eA_\mu) - m)\psi &=& 0, \\
  \bar\psi\bigl(\gamma^\mu(i\overleftarrow{\pd}_\mu + eA_\mu) + m \bigr) &=& 0,
\end{eqnarray}
where $j^\nu = e\bar\psi\gamma^\nu\psi$ is the fermion current, and the left arrow over $\pd_\mu$ indicates that
it acts upon $\bar\psi$, i.e., to the left.

As it can be seen from \eqref{L_MCS4}, the potential-current interaction has the conventional form $-j^\mu
A_\mu$ typical for the Maxwell electrodynamics. We therefore can resort to the pure-photon sector of the MCS
electrodynamics for our consideration of the \emph{boson} Casimir effect. The corresponding canonical
energy-momentum tensor, which follows from the pure-gauge terms in the Lagrangian \eqref{L_MCS4}, reads
\begin{equation}
    T^{\text{can}}_{\mu\nu} = \pd_\nu A^\lambda (F_{\lambda\mu} + \eta^\alpha
    \epsilon_{\alpha\beta\mu\lambda}A^\beta) - \eta_{\mu\nu} \Lagr,
\end{equation}
where $\eta_{\mu\nu} = \mathtt{diag}(1,-1,-1,-1)$ is the Minkowsky spacetime
metric. Unlike the conventional ($\eta^\mu = 0$) case, the canonical
energy-momentum tensor \emph{cannot be ultimately symmetrized}
\cite{ColladayVAK:SME}, so we present it in the following form:
\begin{eqnarray}
  T^{\text{can}}_{\mu\nu} &=& \theta_{\mu\nu} - \pd^\lambda \Xi_{[\lambda\mu]\nu},\\
  \Xi_{[\lambda\mu]\nu} &=& A_\nu(F_{\mu\lambda} - \epsilon_{\alpha\beta\mu\lambda} \eta^\alpha A^\beta),\\
  \theta_{\mu\nu} &=& F_{\mu\lambda}F^\lambda_{\hspace{2pt}\cdot\hspace{1pt}\nu} + \frac14 \eta_{\mu\nu}F_{\alpha\beta}F^{\alpha\beta}
      + \eta^\alpha \epsilon_{\alpha\beta\sigma\rho} A^\gamma F^{\lambda\rho}
                                    \left(
                                                 \delta^\beta_\gamma\delta^\sigma_\mu\eta_{\nu\lambda}
                                                 +\frac12 \delta^\sigma_\lambda
                                                 (\delta^\beta_\mu\eta_{\gamma\nu} -
                                                 \delta^\beta_\gamma\eta_{\mu\nu})
                                    \right),
\end{eqnarray}
where $\Xi_{[\lambda\mu]\nu} = - \Xi_{[\mu\lambda]\nu}$. $\theta_{\mu\nu}$
becomes symmetric and gauge-invariant in the $\eta^\mu = 0$ case.

Like in the conventional electrodynamics, the gauge freedom \eqref{gauge_A}
allows us to choose the transversal gauge (see Ref.~\cite{ColladayVAK:SME})
\begin{equation}
  \pd_\mu A^\mu = 0. \label{4trans_gauge}
\end{equation}
However, in the $\eta^\mu \ne 0$ case, it becomes incompatible with the axial
gauge $n_\mu A^\mu = 0$ ($n_\mu$ is an \textit{arbitrary} constant 4-vector)
\cite{Andrianov}. The only axial gauge that can be fixed together with
\eqref{4trans_gauge} is
\begin{equation}
  \eta_\mu A^\mu = 0    \label{4axial_gauge}.
\end{equation}
A specific feature of the MCS electrodynamics is that both
$T^{\text{can}}_{\mu\nu}$ and even $\theta_{\mu\nu}$ are gauge-dependent
\cite{ColladayVAK:SME}. However, the Casimir force, which is determined
through the normal-normal component of $T^{\text{can}}_{\mu\nu}$ integrated
over the plates, turns out to be gauge-invariant, for the special choice of
$\eta^\mu$ we are interested in. We will prove this statement in the following
subsection.
\subsection{Between the conducting plates}
Here and further on in the paper, we will assume that
\begin{equation}
  \eta^\mu = ( \eta, \bvec{0} ). \label{assumed_eta}
\end{equation}
We will consider the MCS electrodynamics of the free gauge field (without
sources) between the two parallel plates with infinite conductivity. Let us
choose the coordinate system in such a way that the plates are orthogonal to
the $z$ direction (i.e., to the 3rd axis, in accordance with $x^\mu =
(t,x,y,z)$) and correspond to $z = \pm a$, where $D = 2a$ is the distance
between the plates. We will also assume that the plates have the form of large
squares denoted $P_\pm = \{-L/2 \le x,y \le L/2, z = \pm a\}$, with linear
dimensions $L \to \infty$, and that the potential $A_\mu(x)$ satisfies the
periodic boundary conditions,
\begin{equation}\label{A_xy_bcond}
    A_\mu(t, x + L, y, z) = A_\mu(t, x, y + L, z) = A_\mu(t, x, y, z).
\end{equation}
It can be easily shown, using the method analogous to that adopted in the
Maxwell electrodynamics \cite{QEDBoundCond}, that the perfect plate
conductivity implies the same boundary conditions for the electromagnetic
field, as they are in the Maxwell electrodynamics, namely,
\begin{equation}\label{E_z_bcond}
    E_x = E_y = 0, \quad \bvec{x} \in P_\pm,
\end{equation}
where $E^i = F_{0i}$, $i = 1,2,3$, is the electric field strength (we will use
the notation $\bvec{E} = (E_x, E_y, E_z) = (E^1,E^2,E^3)$, $E_i = -E^i$ for
three-dimensional vectors and denote their indices with latin letters
$i,j,k,...$). The Bianchi identity
\begin{equation}\label{Binachy_F}
    \pd_\mu \tilde{F}^{\mu\nu} = 0
\end{equation}
implies that
\begin{eqnarray}
  \pd_0\bvec{H} &=& -\rot\bvec{E},\\
  \divg \bvec{H} &=& 0,   \label{div_H}
\end{eqnarray}
where $\bvec{H}$ is the magnetic field strength, $H^i = -\frac12
\epsilon_{ijk} F_{jk}$ . First of these equations, taken on the plates, gives
\begin{equation}
    \pd_0H_z = \pd_y E_x - \pd_x E_ y = 0, \quad \bvec{x} \in P_\pm. \label{H_z_bcond}
\end{equation}
For the calculation of the Casimir effect, we can not consider the states in which the magnetic field has the
static $z$-component that does not vanish on the plates (the corresponding one-photon eigenstates should have
zero frequencies, but they, as we will show in \eqref{E_vac}, do not contribute to the Casimir energy). Another
thing illustrating this is the fact that zero-frequency Fourier components of $\bvec{E}$ and $\bvec{H}$ are
decoupled in the free MCS equations, so the zero-frequency component of the magnetic field is not constrained
with the boundary conditions \eqref{E_z_bcond}. This means that this mode does not ``feel'' the plates and
therefore does not contribute to the Casimir effect. As a result, we can assume that
\begin{equation}
    H_z = 0, \quad  \bvec{x} \in P_\pm.
\end{equation}
Now it is seen that, for our choice of $\eta^\mu$, the change of the
pure-gauge action \emph{between the plates} under the gauge transformation
\eqref{gauge_A} vanishes:
\begin{eqnarray}
    \delta\mathcal{A} &=& \int d^4x \, \delta\Lagr
       = - \frac{\eta}{2}\int\limits_{-\infty}^{+\infty}dt \int\limits_V dxdydz\,
             \pd_i\left(\alpha  \epsilon_{ijk}F_{jk}\right)\nonumber\\
       &=& \frac{\eta}{2}\int\limits_{-\infty}^{+\infty}dt\int\limits_{-L/2}^{+L/2}dxdy \,
       \alpha(t,x,y,z)H_z(t,x,y,z)\vert_{z=-a}^{z = +a} = 0,
\end{eqnarray}
where $V = \{ |x|, |y| < L/2, |z| < a \}$, i.e., the boundary conditions we
chose do not violate gauge invariance. Of course, in the above expression, the
function $\alpha(x)$ should be taken $x,y$-periodic as the potential
$A_\mu(x)$ itself.

The expression for the energy-momentum tensor becomes simpler when $\eta^\mu$
contains only its timelike component. Of interest are the two following
components:
\begin{eqnarray}\label{T00_special_choice_of eta}
    T^{\text{can}}_{00} &=& \frac12(\bvec{E}^2 + \bvec{H}^2) -\eta \bvec{A}\cdot\bvec{H} + E^i \pd_i A^0,\\
    T^{\text{can}}_{33} &=& \frac12(E_x^2 + E_y^2 - E_z^2 + H_x^2 + H_y^2 - H_z^2) + \Delta T^{\text{can}}_{33}
                                                                         \label{T33_special_choice_of eta} \\
    \Delta T^{\text{can}}_{33} &=&  \pd_\mu(A^3 F^{\mu 3}) + \eta (\epsilon_{3ij} A^i\pd_3A^j
                                    + \bvec{A}\cdot\bvec{H} - 2 A^3 H^3).
\end{eqnarray}
The force acting upon the plates is a result of integration of $T^{\text{can}}_{33}$ over them, i.e., over $x,y
\in [-L/2,L/2]$, and subsequent averaging over the electromagnetic vacuum. To show that this force is
gauge-invariant, let us prove that $\Delta T^{\text{can}}_{33}$ vanishes after this integration and averaging.
Indeed, one can easily show that
\begin{equation}
    \int\limits_{P_\pm} dxdy \, \Delta T^{\text{can}}_{33} =
             \pd_0 \int\limits_{P_\pm} dxdy \, A^3 F^{03}
             + \int\limits_{P_\pm} dxdy \left(\sum\limits_{i=1,2}  \pd_i(A^3 F^{i3})
             -\eta\bigl(\pd_1(A^2 A^3) - \pd_2(A^1 A^3)\bigr)\right).
\end{equation}
The vacuum expectation value of the first term, which is a time derivative, is zero, while the second one
vanishes after the integration over $x, y$ due to the periodicity conditions \eqref{A_xy_bcond}.

Hereby, we have proved that the Casimir force is gauge-invariant. To find the key expression for it, we can now
resort to the axial-transversal gauge \eqref{4axial_gauge}, \eqref{4trans_gauge}, which, for our choice of
$\eta^\mu$, takes the following form:
\begin{eqnarray}
  A^0 &=& 0,             \label{A0_eq_0} \\
  \divg\bvec{A} &=& 0.    \label{3_trans_gauge}
\end{eqnarray}
Within this gauge, the zero-zero component of the energy-momentum tensor reads
\begin{equation} \label{T_00}
  T^{\text{can}}_{00} = \theta_{00} = \frac12(\bvec{E}^2 + \bvec{H}^2) - \eta \bvec{A} \cdot \bvec{H}.
\end{equation}
The equations of motion, in turn, take the form
\begin{equation}
    \square \bvec{A} = 2\eta \rot\bvec{A}. \label{MCS_eqmotion_A_spec_gauge}
\end{equation}
Within this gauge, consider a complete set of one-particle electromagnetic
field eigenstates. These states can be described by potential functions
$\bvec{A} = \bvec{A}_n(\bvec{x}) e^{-i \omega_n t}$ satisfying the
transversality condition \eqref{3_trans_gauge} and the equations of motion
\eqref{MCS_eqmotion_A_spec_gauge}, which take the following form for these
eigenfunctions:
\begin{equation}
        (\omega_n^2 + \nabla^2)\bvec{A}_n = -2\eta \rot\bvec{A}_n, \quad \bvec{x} \in V. \label{secular_main}
\end{equation}
These functions should also be periodic in the $x, y$ directions
\eqref{A_xy_bcond} and satisfy the boundary conditions \eqref{E_z_bcond},
which imply that
\begin{equation} \label{A_z_bcond}
    (A_n)_x = (A_n)_y = 0,  \quad \bvec{x} \in P_\pm,
\end{equation}
since we are not interested in the solutions with $\omega_n = 0$ which do not
contribute to the Casimir effect. Finally, the eigenstates must be normalized
in such a way that
\begin{equation}
        \int_V d^3x \, \bvec{A}_n\cc(\bvec{x}) \cdot \bvec{A}_{n'}(\bvec{x}) = \delta_{n,n'},  \label{A_norm_cond}
\end{equation}
where $n, n'$ are the full sets of quantum numbers (excluding the energy sign
quantum number), which run through a discrete set since we assume the linear
size of the plates $L$ to be finite.

As it can be seen from \eqref{secular_main}, the equations of motion are
time-reversible, since for every solution of the form
$\bvec{A}_n(\bvec{x})e^{-i\omega_n t}$, they also have a solution
$\bvec{A}_n(\bvec{x})e^{+i\omega_n t}$. This comes from the fact that
$\eta^\mu$ is a pseudovector, so its timelike component $\eta$ is invariant
under time reversal.

At the tree level, the electromagnetic potential between the plates can be
quantized in such a way that
\begin{equation} \label{photon_exp}
  \hat{\bvec{A}}(\bvec{x},t) = \sum\limits_n{\frac{1}{\sqrt{2\omega_n}}
                                   \left\{
                                           \bvec{A}_n(\bvec{x})e^{-i\omega_n t}\hat{a}_n +
                                           \bvec{A}_n\cc(\bvec{x})e^{i\omega_n t} \hat{a}_n\hc
                                   \right\}
                                }.
\end{equation}
Operators $\hat{a}\hc_n$/$\hat{a}_n$ commute in the usual way,
\begin{equation}
   [\hat{a}_n, \hat{a}\hc_{n'}] = \delta_{n,n'}, \quad [\hat{a}_n, \hat{a}_{n'}] = [\hat{a}\hc_n,
   \hat{a}\hc_{n'}] = 0,
\end{equation}
and create/destroy one photon in the mode with quantum number $n$. In
Sec.~\ref{sec:SpectrumNotes}, we will show that, for sufficiently small
$|\eta|a$, the imaginary-energy solutions (tachyon modes) form a zero-measure
set, so the processes involving them are very rare, and the quantization
\eqref{photon_exp} is correct.

To find the Casimir force $f_{\text{Casimir}}$ acting upon a unit plate square
(which is gauge-invariant, as it was shown above), we will differentiate the
vacuum energy $E_{\text{vac}}$ with respect to the distance between the
plates,
\begin{eqnarray}
   f_{\text{Casimir}} &=& -\frac{1}{L^2} \frac{\pd E_{\text{vac}}}{\pd D},\\
   E_{\text{vac}}     &=& \brac{0} \int\limits_V d^3x \, T^{\text{can}}_{00}(x) \cket{0}.
\end{eqnarray}
Here, as usual, the negative sign of the force corresponds to the attraction of the plates. Let us show that the
vacuum energy is \emph{half the sum over all positive one-particle eigenstate energies} (as it is in the Maxwell
electrodynamics). Note that for every solution $\bvec{A}(x)$ of the equations of motion
\eqref{MCS_eqmotion_A_spec_gauge}, due to the transversality \eqref{3_trans_gauge} of the potential,
\begin{eqnarray}
   \frac{1}{2}\bvec{H}^2 - \eta \bvec{A}\bvec{H}
   &=& \frac{\bvec{A}(\bvec\nabla\times(\bvec\nabla \times \bvec{A})) + \bvec\nabla(\bvec{A}\times(\bvec\nabla\times\bvec{A}))}{2}
   - \eta \bvec{A}\bvec{H} \nonumber\\
   &=& \frac{\bvec{A}(\bvec\nabla(\bvec\nabla\bvec{A}) - \nabla^2\bvec{A})
   + \bvec\nabla(\bvec{A}\times(\bvec\nabla\times\bvec{A}))}{2} - \eta \bvec{A}\bvec{H}
   = \frac{-\bvec{A}\pd_0^2\bvec{A} + \bvec\nabla(\bvec{A}\times(\bvec\nabla\times\bvec{A}))}{2},
\end{eqnarray}
and, after the integration over $V$, the last term, which is a spatial
divergence, vanishes due to the boundary conditions \eqref{A_z_bcond}. This
implies that
\begin{eqnarray}
    E_{\text{vac}} &=& \brac{0} \int\limits_V d^3x \, T^{\text{can}}_{00}(x) \cket{0}
                   = \brac{0} \int\limits_V d^3x \left(\frac{\bvec{E}^2+\bvec{H}^2}{2} - \eta \bvec{A}\bvec{H}\right) \cket{0}
                   = \brac{0} \int\limits_V d^3x \, \frac{(\pd_0\bvec{A})^2 - \bvec{A}\pd_0^2\bvec{A}}{2} \cket{0}
                   \nonumber\\
                   &=& \sum\limits_n \int\limits_V d^3x \, \frac{1}{2\omega_n} \frac12\left\{
                                         (-i\omega_n\bvec{A}_n)(i\omega_n\bvec{A}_n\cc)
                                          - \bvec{A}_n (i\omega_n)^2\bvec{A}_n\cc\right\}
                   = \sum\limits_n \int\limits_V d^3x \, \frac{\omega_n}{2} \bvec{A}_n\bvec{A}_n\cc
                   = \frac12\sum\limits_n \omega_n, \label{E_vac}
\end{eqnarray}
where we have used the photon decomposition \eqref{photon_exp} and the
normalization condition \eqref{A_norm_cond}. Finally,
\begin{equation}
    f_{\text{Casimir}} = -\frac{1}{L^2} \frac{\pd E_{\text{vac}}(D)}{\pd D}
    = -\frac{1}{L^2} \frac{\pd}{\pd D} \sum\limits_n \frac{\omega_n(D)}{2}.
\end{equation}
The series under the derivative operator is obviously divergent, so it needs
to be regularized and renormalized. One possible way to do this is to
introduce the \emph{zeta function} corresponding to the energy-squared
operator $\hat{H}^2$ that can be extracted from \eqref{secular_main} and has
the eigenvalues equal to $\omega_n^2$:
\begin{eqnarray}
    E_{\text{vac}}^{\text{reg}} &=& \frac12 \zeta_{\hat{H}^2}(-1/2),\\
    \zeta_{\hat{H}^2}(s) &=& \sum_n (\omega_n^2)^{-s},    \label{zeta_Hsquared}\\
    \hat{H}^2 &=& -\nabla^2 - 2\eta \rot.            \label{H_squared}
\end{eqnarray}
The value $\zeta_{\hat{H}^2}(-1/2)$ is the analytical continuation of the
series \eqref{zeta_Hsquared}, which is convergent for large $\RRe s$, $s \in
\Complex$.

Another possible renormalization method involves subtraction of an expression of the form $(C^{(1)} + C^{(2)}
D)$, where $C^{(1,2)}$ are infinite constants, from the vacuum energy. $C^{(1)}$ is obviously nonphysical, while
$C^{(2)}$ equals the force acting upon the external side of the plate, i.e., the Casimir force in a semispace,
so subtracting it is also physically motivated (see page \pageref{page:physRenorm}). Zeta function
regularization automatically renormalizes the vacuum energy.

We will use these two techniques in sections \ref{sec:ZetaReg},
\ref{sec:ResidueTheorem}, respectively, to find the corrections to the
electromagnetic Casimir force due to the existence of a nonzero $\eta$
coupling.
\section{Electromagnetic field modes between the plates}\label{sec:PhotonEigenStates}
In this section, we obtain the expressions for the solutions
$\bvec{A}_n(\bvec{x})$ of vacuum Maxwell-Chern-Simons secular equation
\eqref{secular_main} with $\eta^\mu = (\eta, \bvec{0} )$, i.e., solve the free
photon eigenstate problem.

The translational invariance along the $x, y$ directions allows us to search
for the solutions having the form
\begin{equation}
    \bvec{A}_n(\bvec{x}) = N e^{i \bvec{k} \bvec{x}} \bvec{f}(z), \qquad \bvec{k} = (k_x, k_y, 0 ),
\end{equation}
where $N$ is the normalization constant. Due to the invariance with respect to
rotations on the $z$ axis, we can choose the cartesian coordinate system so
that $k_y = 0$, $k_x = k \ge 0$. Of course, the periodicity conditions
\eqref{A_xy_bcond} with $L \to \infty$, which are not invariant under these
rotations, do not affect the form of the solutions, but only the normalization
coefficient $N$ and the possible values of $\bvec{k}$, so that $k_{x,y} =
\frac{2\pi}{L}n_{x,y}$, $n_{x,y} \in \Ints$.

The gauge-fixing condition \eqref{3_trans_gauge} takes the form
\begin{equation}
    k f_x = i \pd_z f_z.    \label{gauge_xyz}
\end{equation}
On the boundary (i.e., plates, $z = \pm a$), $A_{x,y} = 0$, hence
\begin{eqnarray}
   f_{x,y}(\pm a) &=& 0, \label{bnd_cond_xy}\\
   \pd_z f_z(\pm a) &=& 0. \label{bnd_cond_z}
\end{eqnarray}
The projections of the MCS equations of motion \eqref{secular_main} onto the
$x, y, z$ axes read, respectively,
\begin{eqnarray}
  (k^2 - \omega_n^2 - \pd_z^2)f_x &=& -2\eta \pd_z f_y,               \label{sec_main_x} \\
  (k^2 - \omega_n^2 - \pd_z^2)f_y &=& 2\eta (-i k f_z + \pd_z f_x),  \label{sec_main_y} \\
  (k^2 - \omega_n^2 - \pd_z^2)f_z &=& 2\eta \cdot i k f_y,          \label{sec_main_z}
\end{eqnarray}
where $\omega_n$ is the energy corresponding to the eigenstate. Note that
differentiation of \eqref{sec_main_z} with respect to $z$ and subsequent
substitution $\pd_z f_z = - i k f_x$ (according to \eqref{gauge_xyz}) gives
\eqref{sec_main_x}, so we can discard the latter equation from our
consideration.

Using \eqref{gauge_xyz}, substitute $\frac{i}{k} \pd_z^2 f_z$ for $\pd_z f_x$
in \eqref{sec_main_y}, then we come to the system of equations on $f_y$,
$f_z$:
\begin{eqnarray}
  (k^2 - \omega_n^2 - \pd_z^2)k f_y &=& -2i\eta (k^2 - \pd_z^2) f_z,  \label{sec_main_y_2} \\
  (k^2 - \omega_n^2 - \pd_z^2)f_z &=& 2i\eta k f_y.          \label{sec_main_z_2}
\end{eqnarray}
These equations, together with the boundary and gauge-fixing conditions,
possess a special kind of discrete symmetry (we will call it the $z$-parity
and denote $\hat\Pi$):
\begin{gather}
  \hat\Pi A_i(x,y,z) = \left(\delta_{ij} - \frac{2\hat{k}_i \hat{k}_j}{\hat{\bvec{k}}^2}\right)A_j(x,y,-z),
  \qquad \hat{\bvec{k}} = ( -i \pd_x, -i\pd_y, 0),\\
  \hat\Pi f_{y,z}(z) = f_{y,z}(-z), \quad \hat\Pi f_x(z)      = -f_x(-z)   \label{Pi_yz}.
\end{gather}
The transformation law of the $x$-component, compared to that of the
$y,z$-components, is in accordance with \eqref{gauge_xyz}. Let
$\bvec{A}(\bvec{x})$ (and $\bvec{f}(z)$, in turn) be the eigenfunction of the
$z$-parity with the eigenvalue $\Pi$, which is obviously $+1$ or $-1$, since
$\hat\Pi^2 = \hat1$.

The solution $\bvec{f}(z)$ corresponding to definite $\Pi$ and $k$ is a
superposition of the two solutions $\bar{\bvec{f}}^{(\Pi,\lambda)}(z)$ in the
boundless space (i.e., without the boundary conditions) with the two
polarizations $\lambda = \pm1$. As it is seen from \eqref{sec_main_y_2} and
\eqref{sec_main_z_2}, we can search for the $y,z$-components of these
solutions in the form
\begin{eqnarray}
    \bar{f}^{(\Pi,\lambda)}_{y,z}(z) &\propto& \varphi_\Pi(\varkappa z),\\
    \varphi_\Pi(\varkappa z) &\equiv& \twoCases{\cos{\varkappa z}}{\Pi = +1}{\sin{\varkappa z}}{\Pi = -1,}
\end{eqnarray}
where $\varkappa$ is a complex constant. Then the equations
\eqref{sec_main_y_2}, \eqref{sec_main_z_2} are easily solved, and we obtain
\begin{eqnarray}
  \twoCol{\bar{f}^{(\Pi,\lambda)}_y}{\bar{f}^{(\Pi,\lambda)}_z} &=&
  \twoCol{1}{-i \lambda \cos\theta_\lambda} \varphi_\Pi(\varkappa_\lambda z),  \quad \lambda,\Pi = \pm1, \label{boundless_f_yz}\\
  K_\lambda^2 &=& -\eta\lambda + \sqrt{\omega_n^2 + \eta^2},  \label{vac_spec}\\
  \varkappa_\lambda &=& \sqrt{K_\lambda^2 - k^2}, \\
  \cos\theta_\lambda &=& \frac{k}{K_\lambda}, \quad \sin\theta_\lambda = \frac{\varkappa_\lambda}{K_\lambda}.
\end{eqnarray}
As seen, the sign of $\varkappa_\lambda$ does not affect the form of the solution, except for its sign. Note
that the solutions with complex $\varkappa_\lambda$, which are unbounded for $|z| \to \infty$, should not be
discarded from consideration when finding the solutions \emph{between the plates}, where $z$ is finite. In the
next section (see page \pageref{kappa_only_img_real}), we will show that $\varkappa_\lambda$ can be either real
or pure imaginary for these solutions.

Now consider a solution of \eqref{sec_main_y_2}, \eqref{sec_main_z_2} with definite $\Pi = \pm1$ satisfying the
boundary conditions \eqref{bnd_cond_xy}, \eqref{bnd_cond_z}
\begin{equation}
    f_{y,z}(z) = \sum\limits_{\lambda = \pm1} C_\lambda \bar{f}^{(\Pi,\lambda)}_{y,z}(z).
\end{equation}
Since the boundary conditions at $z = +a$ and $z = -a$ are equivalent for the
functions with definite $z$-parity $\Pi$, it is enough to consider these
conditions only at the point $z = +a$. In the following expressions, we denote
$\xi_{\pm} \equiv \xi_{\pm1}$ for the symbols depending on the polarization
$\lambda$, such as $C_\lambda$, $\varkappa_\lambda$, $\theta_\lambda$, etc.
Then the boundary conditions for $f_y$ and $f_z$ take the form:
\begin{align}
   C_+ \varphi_\Pi(\varkappa_+ a) &+ C_- \varphi_\Pi(\varkappa_- a) &= 0,    \label{C_system_1}\\
   C_+ \cdot (-i) \cos\theta_+ (-\Pi \varkappa_+ \varphi_{-\Pi}(\varkappa_+ a)) &+
   C_- \cdot i \cos\theta_-  (-\Pi \varkappa_- \varphi_{-\Pi}(\varkappa_- a)) &= 0,   \label{C_system_2}
\end{align}
where we have used the identity
\begin{equation}
    \pd_z \varphi_\Pi(\varkappa z) = -\Pi \varkappa \varphi_{-\Pi}(\varkappa z).
\end{equation}
The boundary conditions for $f_x$ are satisfied in turn, due to
\eqref{gauge_xyz}. The existence of nontrivial solutions for $C_\pm$ requires
that the determinant of the system of equations \eqref{C_system_1},
\eqref{C_system_2} vanishes, then
\begin{equation}
    g_\Pi(\omega_n^2) \equiv  \varphi_\Pi(\varkappa_+ a) \varphi_{-\Pi} (\varkappa_- a)  \sin\theta_- +
                            \varphi_\Pi(\varkappa_- a) \varphi_{-\Pi} (\varkappa_+ a)  \sin\theta_+ = 0.  \label{spec_cond}
\end{equation}
This equation implicitly determines the spectrum for every $k \ge 0$ and $\Pi
= \pm1$. When it is solved, the solutions for $C_\pm$ read (up to a nonzero
multiplicative constant)
\begin{equation}
    C_\pm = \pm \varphi_\Pi(\varkappa_\mp a).  \label{C_pm}
\end{equation}
Now, to determine the final expressions for $\bvec{f}(z)$, we use the
solutions without the plates \eqref{boundless_f_yz} and then
Eq.~\eqref{gauge_xyz} to find $f_x(z)$. As a result, we obtain
\begin{equation} \label{vec_f_solution}
   \begin{array}{rcrcl}
      \bvec{f}(z) &=& - \bvec{e}_x &\cdot& \Pi (\sin\theta_+\varphi_\Pi(\varkappa_- a) \varphi_{-\Pi}(\varkappa_+ z) +
                                           \sin\theta_-\varphi_\Pi(\varkappa_+ a) \varphi_{-\Pi}(\varkappa_- z))  + \\
                  & & + \bvec{e}_y &\cdot&     (\varphi_\Pi(\varkappa_- a) \varphi_{\Pi}(\varkappa_+ z) -
                                             \varphi_\Pi(\varkappa_+ a) \varphi_{\Pi}(\varkappa_- z)) - \\
                  & & - \bvec{e}_z &\cdot&  i  (\cos\theta_+\varphi_\Pi(\varkappa_- a) \varphi_{\Pi}(\varkappa_+ z) +
                                           \cos\theta_-\varphi_\Pi(\varkappa_+ a) \varphi_{\Pi}(\varkappa_- z)),
   \end{array}
\end{equation}
where, for arbitrary $\bvec{k} = (k_x, k_y, 0)$,
\begin{equation}
    \bvec{e}_x = \frac{\bvec{k}}{|\bvec{k}|}, \quad \bvec{e}_y = \frac{1}{|\bvec{k}|} \bvec{e}_z\times\bvec{k},
\end{equation}
and $\bvec{e}_z$ is a normal to the plates pointing in the direction of
positive $z$, i.e., $\bvec{e}_z = \bvec\nabla z$. The full set of quantum
numbers is $n = (k_x, k_y, \Pi, m)$, where $k_x = \frac{2\pi n_x}{L}$, $k_y =
\frac{2\pi n_y}{L}$, $n_{x,y} \in \Ints$, $\Pi = \pm1$, and the last quantum
number $m = 0,1,2,3,...$ enumerates the solutions of the spectrum equation
\eqref{spec_cond}.

When $\eta = 0$, i.e., in the Maxwell case, \eqref{spec_cond} gives the
well-known spectrum
\begin{eqnarray}
   \varkappa_+, \varkappa_- &\to& \varkappa_{\text{M}} = \frac{\pi m}{2a}, \\
   \omega_n                 &\to& (\omega_n)_{\text{M}} = \sqrt{k^2 + \left(\frac{\pi m}{2a}\right)^2},
                                                       \quad m = 0,1,2,3. \label{omega_Maxwell}
\end{eqnarray}
The solution for the potential, corresponding to $m = 0$ and $\Pi = -1$ is
trivial in this case, the others are physical. However, the energy eigenvalue
with $m = 0, \Pi = +1$ does not depend on $a$ and hence does not contribute to
the Casimir energy.
\section{General notes about the energy spectrum}\label{sec:SpectrumNotes}
In this section, we will show that the Maxwell-Chern-Simons photon energy
spectrum between the plates is such that the energy-squared eigenvalues
$\omega_n^2$ are real, $\omega_n^2 \ge -{\eta}^2$, and, when $|\eta| <
\frac{\pi}{4a}$, the imaginary-energy solutions have $\omega_n^2 = -{\eta}^2$
and form a zero-measure set in the quantum number space, i.e., almost no
tachyon modes are present.

Consider a space $\mathcal{F}$ of potential functions $\bvec{A}(\bvec{x})$,
defined in a spatial domain $V$ between the plates, which are transversal
\eqref{3_trans_gauge} and $x,y$-periodic \eqref{A_xy_bcond}. The scalar
product and the norm in $\mathcal{F}$ are defined in the usual way,
\begin{equation}
    \bracket{A'}{A} \equiv \int\limits_V d^3x \, (\bvec{A}'(\bvec{x}))\cc \bvec{A}(\bvec{x}),
    \quad \|A\| \equiv \sqrt{\bracket{A}{A}}.
\end{equation}
It can be easily seen that the energy-squared operator $\hat{H}^2$ (see
Eq.~\eqref{H_squared}) acts inside of $\mathcal{F}$, since
$\forall\bvec{A}\in\mathcal{F}: \divg(\hat{H}^2\bvec{A}(\bvec{x})) = 0$ .
Denote the orthonormalized eigenstates of $\hat{H}^2$ in $\mathcal{F}$ (which
were discussed in the previous section) as $\cket{A_n}$, so that
$\hat{H}^2\cket{A_n} = \omega_n^2 \cket{A_n}$. The corresponding
eigenfunctions $\bvec{A}_n(\bvec{x})$ should also satisfy the boundary
conditions \eqref{A_z_bcond} on the plates $P_\pm$, which, in turn, imply that
$\pd_z(A_n)_z = \divg\bvec{A}_n - \pd_x(A_n)_x - \pd_y(A_n)_y = 0$ on them.
Let us prove the following assertion.
\begin{assertion} \label{ass:1}
  The eigenvalues $\omega_n^2$ are real and $\omega_n^2 \ge -\eta^2$.
\begin{proof}
  Let us denote $\xi_s = \brac{A_n}\rot\cket{A_n}$ and $\xi_p = \brac{A_n}(-\nabla^2)\cket{A_n}$, then
  $\omega_n^2 = \brac{A_n}\hat{H}^2\cket{A_n} = \brac{A_n} -\nabla^2 -2\eta\rot \cket{A_n} = \xi_p - 2\eta \xi_s$.
  First, integrating by parts, we obtain
  \begin{equation}
     \xi_p = \int\limits_V (\bvec{A}_n)_i\cc (-\pd_j\pd_j (\bvec{A}_n)_i) d^3x
           = \int\limits_V |\pd_j (A_n)_i|^2 d^3x - \int\limits_{P} (A_n)_i\cc \pd_z (A_n)_i dS_z,
     \quad P = P_+ \cup P_-,
  \end{equation}
  where the area element $dS_z = \pm dS = \pm |dxdy|$ on $P_\pm$. Indeed, due to the periodicity,
  the integration by parts over $x, y$ does not give the surface term analogous to the last one in the above expression.
  The latter term, however, vanishes due to the boundary conditions on $P_\pm$, namely,
  $(A_n)_x = (A_n)_y = 0$, $\pd_z(A_n)_z = 0$.
  Thus,
  \begin{equation}
    \xi_p = \int\limits_V |\pd_j (A_n)_i|^2 d^3x \in [0,+\infty), \label{xi_p}
  \end{equation}
  i.e., real and nonnegative. Making the analogous integration by parts in the expression for $\xi_s$, we obtain
  \begin{equation}
    \xi_s = \int\limits_V \epsilon_{ijk} (A_n)_i\cc \pd_j (A_n)_k d^3x =
    \int\limits_V (-\pd_j (A_n)_i\cc)  \epsilon_{ijk} (A_n)_k  d^3x
    + \int\limits_P \epsilon_{i3k} (A_n)_i\cc (A_n)_k dS_z,
  \end{equation}
  and $\epsilon_{i3k} (A_n)_i\cc (A_n)_k = (A_n)_y\cc (A_n)_x - (A_n)_x\cc (A_n)_y = 0$ on $P_\pm$. Finally,
  \begin{equation}
    \xi_s =     \int\limits_V (-\pd_j (A_n)_i\cc) \epsilon_{ijk} (A_n)_k  d^3x =
                \int\limits_V (A_n)_k \epsilon_{kji} (\pd_j (A_n)_i)\cc = \xi_s\cc,
  \end{equation}
  meaning that $\xi_s$ is real. Thus we have proved that $\omega_n^2 = \xi_p - 2\eta \xi_s$ is real.

  Now let us show that $\omega_n^2 \ge -{\eta}^2$. Consider the two states $\cket{A'} = \cket{A_n}$
  and $\cket{A} = \rot \cket{A_n}$. Note that $\cket{A} \in \mathcal{F}$. Then, according to the Cauchy-Bunyakovsky-Schwarz inequality
  \cite{CauchyBunyakovskySchwarz},
  $|\bracket{A'}{A}|^2 \le \|A'\|^2  \|A\|^2$, i.e.,
  \begin{equation}
  \xi_s^2  =  |\brac{A_n}\rot\cket{A_n}|^2 \le  \| \rot \cket{A_n} \|^2
  = \int\limits_V (A_n)_j\cc \epsilon_{iaj} \overleftarrow{\pd}_a \epsilon_{ibk} \pd_b (A_n)_k  d^3x =
    \int\limits_V (A_n)_j\cc (\overleftarrow{\pd}_a \delta_{jk} \pd_a - \overleftarrow{\pd}_k \pd_j) (A_n)_k
    d^3x.
  \end{equation}
  Now, the integration of the second term by parts makes $\overleftarrow{\pd}$ operators act to the right, and we obtain
  \begin{equation}
    \xi_s^2 \le \int\limits_V |\pd_a(A_n)_k|^2 d^3x + \int\limits_V (A_n)_j\cc \pd_k\pd_j(A_n)_k d^3x
              -\int\limits_{P} (A_n)_j\cc \pd_j (A_n)_z dS_z,
  \end{equation}
  where the second term vanishes due to the transversality, while the third due to the boundary conditions. Finally,
  comparing with \eqref{xi_p}, we can conclude that
  \begin{equation}
    \xi_s^2 \le \xi_p.
  \end{equation}
  Hence, the energy-squared eigenvalue
  \begin{equation}\label{omega_constraint1}
    \omega_n^2 = \xi_p - 2\eta \xi_s \ge \xi_p - 2 |\eta| \sqrt{\xi_p} = (\sqrt{\xi_p} - |\eta|)^2 - {\eta}^2
    \ge -{\eta}^2.
  \end{equation}
\end{proof}
\end{assertion}
As seen, the assertion proved above is also valid in the infinite space ($a
\to \infty$). Then, since $\omega_n^2 + {\eta}^2 \ge 0$, the two branches of
the spatial momentum $K_\lambda$, which are defined in \eqref{vac_spec}, are
real, while $\varkappa_\lambda = \sqrt{K_\lambda^2 - k^2}$ can be either real
or pure imaginary. \label{kappa_only_img_real} Moreover, one can easily see
that
\begin{equation}
    \omega_n^2 = K_+ K_-,
\end{equation}
i.e., the imaginary-energy solutions correspond to $K_\pm$ having opposite
signs, and the real-energy solutions correspond to $K_\pm \ge 0$. We can also
note that $\cos\theta_\lambda = k/K_\lambda$ is real however it may not lie
within the segment $[-1,1]$, and $\sin\theta_\lambda = \varkappa_\lambda /
K_\lambda$ is either real or imaginary (the latter case when
$|\cos\theta_\lambda| > 1$).

Further we show that, for sufficiently small finite $a$, the imaginary-energy
states are negligibly few.
\begin{assertion}
  For the solutions with negative $z$-parity $\Pi = -1$, $\xi_p \ge \frac{\pi^2}{4a^2}$.
  \begin{proof}
     As it was shown in the previous section, we can choose the coordinate system in such a way that
     \begin{equation}
        \bvec{A}_n(\bvec{x}) = N \bvec{f}(z) e^{i k x}, \quad k > 0.
     \end{equation}
     Then, using \eqref{xi_p}, we obtain
     \begin{eqnarray}
        \|A_n\|^2 &=& |N|^2 L^2 \|\bvec{f}\|^2, \label{A_f_norm}\\
        \xi_p \equiv \brac{A_n}(-\nabla^2)\cket{A_n} &=& \int\limits_V |\pd_i(A_n)_j|^2 d^3x =
                      |N|^2 L^2 (k^2 \|\bvec{f}\|^2 + \|\pd_z\bvec{f}\|^2),  \label{nabla2A_f_norm}\\
        \|\bvec{f}\|^2 &\equiv& \int\limits_{-a}^{a} |\bvec{f}(z)|^2 dz, \quad
        \|\pd_z\bvec{f}\|^2 \equiv \int\limits_{-a}^{a} |\pd_z\bvec{f}(z)|^2 dz.
     \end{eqnarray}
     Due to the boundary conditions \eqref{bnd_cond_xy}, \eqref{bnd_cond_z}, the negative $z$-parity
     $\hat{\Pi}\bvec{f}(z) = -\bvec{f}(z)$, and the transversality \eqref{gauge_xyz},
     we can write $\bvec{f}(z)$ as a Fourier series on the segment $z \in [-a,a]$:
     \begin{eqnarray}
       f_x(z) &=&  \frac{i}{k} \sum\limits_{m=1}^\infty\frac{\pi(m-1/2)}{a} \,  \alpha_{z,m} \cos\frac{\pi(m-1/2)z}{a}, \\
       f_y(z) &=&  \sum\limits_{m=1}^\infty \alpha_{y,m} \sin\frac{\pi m z}{a}, \\
       f_z(z) &=&  \sum\limits_{m=1}^\infty \alpha_{z,m} \sin\frac{\pi(m-1/2)z}{a}.
     \end{eqnarray}
     Using the fact that
     \begin{eqnarray}
        \int\limits_{-a}^a \sin\frac{\pi m z}{a} \sin\frac{\pi m' z}{a} dz =
        \int\limits_{-a}^a \cos\frac{\pi m z}{a} \cos\frac{\pi m' z}{a} dz &=& a \delta_{m,m'},\\
        \int\limits_{-a}^a \sin\frac{\pi (m-1/2) z}{a} \sin\frac{\pi (m'-1/2) z}{a} dz &=& a \delta_{m,m'},\\
        \int\limits_{-a}^a \cos\frac{\pi (m-1/2) z}{a} \cos\frac{\pi (m'-1/2) z}{a} dz &=& a \delta_{m,m'},
        \quad m, m' = 1,2,3,...,
     \end{eqnarray}
     we find the expressions for the norms in terms of the $\alpha$-coefficients:
     \begin{eqnarray}
       \|\bvec{f}\|^2 &=& a\sum\limits_{m=1}^\infty \left\{ |\alpha_{y,m}|^2 +
                                                         \left(1 + \left(
                                                                     \frac{\pi(m-1/2)}{k a}
                                                                   \right)^2
                                                         \right) |\alpha_{z,m}|^2
                                                   \right\}, \\
       \|\pd_z\bvec{f}\|^2 &=& a\sum\limits_{m=1}^\infty \left\{ \left|\frac{\pi m}{a} \alpha_{y,m}\right|^2 +
                                                         \left(1 + \left(
                                                                     \frac{\pi(m-1/2)}{k a}
                                                                   \right)^2
                                                         \right) \left| \frac{\pi(m-1/2)}{a}\alpha_{z,m}\right|^2
                                                   \right\}.
    \end{eqnarray}
    Now note that $\left|\frac{\pi m}{a} \alpha_{y,m}\right|^2 \ge \frac{\pi^2}{4a^2}|\alpha_{y,m}|^2$,
    $\left|\frac{\pi (m-1/2)}{a} \alpha_{z,m}\right|^2 \ge \frac{\pi^2}{4a^2}|\alpha_{z,m}|^2$, then the following inequality
    takes place:
    \begin{equation}
        \|\pd_z\bvec{f}\|^2 \ge \frac{\pi^2}{4a^2}\|\bvec{f}\|^2,
    \end{equation}
    which, due to \eqref{A_f_norm}, \eqref{nabla2A_f_norm},
    leads to the lower constraint for $\xi_p$ in a normalized state $\cket{A_n}$ with negative $z$-parity:
    \begin{equation}
        \xi_p = \frac{\brac{A_n}(-\nabla^2)\cket{A_n}}{\|A_n\|^2} = k^2 + \frac{\|\pd_z\bvec{f}\|^2}{\|\bvec{f}\|^2} \ge
        k^2 + \frac{\pi^2}{4a^2} \ge \frac{\pi^2}{4a^2}.
    \end{equation}
  \end{proof}
\end{assertion}
Together with \eqref{omega_constraint1}, this obviously leads to the
statement.
\begin{assertion}
  When $a \le \frac{\pi}{4|\eta|}$, the negative $z$-parity eigenstates correspond to the eigenvalues $\omega_n^2 \ge 0$.
\end{assertion}
\vspace{1em} The method we used above to place the constraint on the energy spectrum of the negative $z$-parity
states does not apply to the states with positive $z$-parity. Indeed, the state with $\bvec{f}(z) = ( 0, 0,
f_z)$, $f_z = const$, which satisfies the gauge and the boundary conditions, corresponds to $\xi_p = k^2$, which
can be arbitrary small nonnegative number. This is inconsistent with the method we have used in the proofs
above. However, it is possible to derive a different spectrum constraint directly analyzing the properties of
the equation $g_+(\omega_n^2) = 0$  (see Eq.~\eqref{spec_cond}), which implicitly determines the energy spectrum
of positive $z$-parity states.

\begin{assertion}
  When $a \le \frac{\pi}{2|\eta|}$, the only nontrivial solution with $\omega_n^2 = -{\eta}^2$ has the form
  $\bvec{f}(z) = ( 0, 0, const)$ for $k = |\eta|$.
  \begin{proof}
    Note that when one replaces $\eta$ with $-\eta$, then $K_+ \leftrightarrow K_-$, and all quantities that depend
    on the polarization $\lambda = \pm1$ flip in the same way. The functions $g_\Pi(\omega^2)$ are symmetric with respect
    to this polarization flip, so they are also invariant under the change of the sign of $\eta$,
    \label{gPi_eta_parity}
    and the properties of
    the energy spectrum, in turn, do not depend on it.
    In view of this, we will assume that $\eta \ge 0$ in our further speculations.

    For $\omega_n^2 = -{\eta}^2$ (minimum possible value), $K_\pm = \mp\eta$,
    $\varkappa_+ = \varkappa_- = \sqrt{{\eta}^2 - k^2}$, and $g_+ = 0$. However, we should also analyze
    the existence of \emph{nontrivial} solutions for $\bvec{f}(z)$, and not only for $C_\pm$ (see
    Eq.~\eqref{C_pm}). Then let us write the expressions for $f_{y,z}(z)$
    for this case (defined in the previous section):
    \begin{equation}
        f_{y,z}(z)  = \alpha_{y,z} \cos\left(\sqrt{{\eta}^2 - k^2} \, z\right), \quad |\alpha_y|^2 + |\alpha_z|^2 > 0.
    \end{equation}
    For $k > \eta$, this solution is inconsistent with the boundary conditions \eqref{bnd_cond_xy}, \eqref{bnd_cond_z}.
    When $k = \eta$, the only solution is the one announced in the assertion,
    \begin{equation} \label{vac_ground_state}
       f_y(z) = 0, \quad f_z(z) = const.
    \end{equation}
    This solution exists for any $a > 0$, so it makes a constant contribution to the vacuum energy, and thus does not
    affect the Casimir force. Moreover, the solutions of this type form a zero-measure set compared to the complete
    set of eigenfunctions, since they exist only for $k = \eta$, while other solutions exist
    within a finite-measure set in the $\bvec{k}$-space. This means that the contribution of the quantum processes involving
    these states is negligible.

    In the $k < \eta$ case, the boundary conditions are satisfied with the following two types of solutions:
    \begin{align}
      f^{(1)}_y(z) &= 0,&    f^{(1)}_z(z) &= const \cdot \cos\frac{\pi m z}{a},& k = \sqrt{{\eta}^2 - (\pi m/a)^2} \ge 0,\\
      f^{(2)}_y(z) &= const \cdot \cos\frac{\pi (m-1/2) z}{a} ,& f^{(2)}_z(z) &= 0,&  k = \sqrt{{\eta}^2 - (\pi (m-1/2)/a)^2} \ge 0,
    \end{align}
    where $m = 1,2,3,...$  When $a < \frac{\pi}{2\eta}$, the solutions are inconsistent with $k \ge 0$ for any $m = 1,2,3,...$,
    so there are no solutions of this type. Therefore, in this case, the only solution with $\omega_n^2 = -\eta^2$ has the form
    \eqref{vac_ground_state}.
  \end{proof}
\end{assertion}
\begin{assertion}
   When $a < \frac{\pi}{4|\eta|}$, the solutions with $-{\eta}^2 < \omega_n^2 < 0$ do not exist.
   \begin{proof}
     Let us introduce the two functions:
     \begin{equation}\label{g_tilde}
        \tilde{g}_\Pi(\omega^2) \equiv \frac{g_\Pi(\omega^2)}{\varphi_\Pi(\varkappa_+ a) \varphi_\Pi(\varkappa_- a)} =
                \tan^\Pi{\varkappa_+ a} \, \sin\theta_+ + \tan^\Pi{\varkappa_- a} \, \sin\theta_-.
     \end{equation}
     Let us treat $g_\Pi$ and $\tilde{g}_\Pi$ as functions of $K_+ \in [-\eta, +\infty)$, with $K_- = K_+ + 2\eta$,
     $\eta \ge 0$ (both $g_\Pi$ and $\tilde{g}_\Pi$ do not depend on the signs of $\eta$ and $\omega$). This interval
     is in the one-to-one correspondence with $\omega^2 = K_+ K_- \in [-\eta^2, +\infty)$. Consider the domain
     with negative $\omega^2 \in (-{\eta}^2, 0)$, which corresponds to $K_+ \in (-\eta, 0)$.
     Here, $\varkappa_\pm \in (+i\infty, 0) \cup [0, 2\eta)$ and, because we assume that $\eta < \frac{\pi}{4a}$,
     $\cos\varkappa_\pm z \ne 0$ for any $z \in [-a,a]$, so the denominator in the expression \eqref{g_tilde}
     for $\tilde{g}_+$ is nonzero. Thus, the equations $g_+ = 0$ and $\tilde{g}_+ = 0$ are equivalent in the domain
     $K_+ \in (-\eta, 0)$, moreover, $\tilde{g}_+$ is a real continuous function of $K_+ \in (-\eta, 0)$ for any $k \ge 0$.
     At the end points of this interval,
     \begin{eqnarray}
       \tilde{g}_+|_{K_+ = -\eta_+} &=& 0,                 \label{gPlus_endPoints1}\\
       \tilde{g}_+|_{K_+ \to 0_-}     &\to& +\infty.           \label{gPlus_endPoints2}
     \end{eqnarray}
     Taking the derivative of $\tilde{g}_+$ with respect to $K_+$, we obtain
     \begin{equation}\label{gPlus_deriv}
           \frac{1}{a}\frac{\pd \tilde{g}_+}{\pd K_+} =  \sum\limits_{\lambda = \pm1}
                                                  \left(
                                                      1 + \tan^2\varkappa_\lambda a
                                                        + \frac{\tan\varkappa_\lambda a}
                                                               {\varkappa_\lambda a}\cos^2\theta_\lambda
                                                  \right).
     \end{equation}
     Here, $\cos^2\theta_\pm = k^2 / K_\pm^2 \ge 0$. On the other hand, if $\varkappa_\lambda$ is imaginary, then
     \begin{eqnarray}
       1 + \tan^2\varkappa_\lambda a &=& 1 - \tanh^2|\varkappa_\lambda a| > 0, \\
       \frac{\tan\varkappa_\lambda a}{\varkappa_\lambda a} &=&
       \frac{\tanh|\varkappa_\lambda a|}{|\varkappa_\lambda a|} > 0.
     \end{eqnarray}
     If $\varkappa_\lambda$ is real, then $0 \le \varkappa_\lambda a < 2\eta a < \pi / 2$,
     hence $\tan\varkappa_\lambda a \ge 0$, and
     \begin{eqnarray}
       1 + \tan^2\varkappa_\lambda a &\ge& 1, \\
       \frac{\tan\varkappa_\lambda a}{\varkappa_\lambda a} &\ge& 0.
     \end{eqnarray}
     Finally, we conclude that the whole expression \eqref{gPlus_deriv} is positive:
     \begin{equation}
           \frac{\pd\tilde{g}_+}{\pd K_+} > 0 \quad  \text{for $K_+ \in (-\eta, 0)$ and $k \ge 0$.}
     \end{equation}
     Taken together with \eqref{gPlus_endPoints1}, \eqref{gPlus_endPoints2}, this proves that $g_+(\omega^2)$ is
     positive within the interval $K_+ \in (-\eta, 0)$, and the corresponding one-photon eigenstates do not exist.
  \end{proof}
\end{assertion}
Thus, we have come to the following conclusion.
\begin{assertion}\label{FinalAssertion}
  When $a < \frac{\pi}{4|\eta|}$, the only negative-$\omega^2$ Maxwell-Chern-Simons one-photon energy states
  are the following:
  \begin{eqnarray}
       \bvec{A}_n(\bvec{x}) &=& N e^{i \bvec{k}\bvec{x}} \bvec{e}_z,\\
       \omega_n^2 &=& -{\eta}^2, \\
       \bvec{k} &=& ( k_x, k_y, 0), \, |\bvec{k}| = |\eta|, \quad \bvec{e}_z = (0,0,1).
  \end{eqnarray}
  These states form a zero-measure set in the $\bvec{k}$-space. All other states are non-tachyonic
  (i.e., correspond to $\omega_n^2 \ge 0$).
\end{assertion}
\section{The Casimir energy: zeta function regularization}\label{sec:ZetaReg}
In this section, we will find the corrections to the Casimir force and the
Casimir vacuum energy calculating the leading $\eta$-correction to the zeta
function defined in \eqref{zeta_Hsquared}, in the case $|\eta| a \ll 1$. As it
was shown in the previous section, in this case, the solutions with imaginary
energies form a zero-measure set in the $\bvec{k}$-space, then the
\emph{reduced zeta function}, defined per unit plate square,
\begin{equation} \label{zeta_red}
    \zeta(s) = \frac{1}{L^2}\zeta_{\hat{H}^2}(s) = \frac{1}{L^2}\sum\limits_n (\omega_n^2)^{-s}
                                                 = \int\limits_0^\infty \frac{k dk}{2\pi}
                                                   \sum\limits_{\Pi=\pm 1}\sum\limits_m (\omega_{k,\Pi,m}^2)^{-s}
\end{equation}
does not include the contribution from these states which vanishes divided by $L^2 \to \infty$. On the other
hand, the contributions from all real-frequency states are unambiguous, since $(\omega_n^2)^{-s} = e^{-s
\log\omega_n^2}$ when $\omega_n^2 > 0$. Here, $n = (k_x, k_y, \Pi, m)$ is a full set of quantum numbers (see
Sec.~\ref{sec:PhotonEigenStates}). The series \eqref{zeta_red} typically converges for sufficiently large $\RRe
s$ and thus can be analytically continued to $s \in \Complex$, in particular, to the point $s = -1/2$, which
corresponds to the renormalized sum over the frequencies $\omega_{k,\Pi,m}$.

As mentioned in the previous section (see page \pageref{gPi_eta_parity}), the
frequencies $\omega_n$ do not depend on the sign of $\eta$, then $\zeta(s)$ is
also an even function of $\eta$, or is, in other words, a function of
${\eta}^2$. Then the derivative of the zeta function with respect to
${\eta}^2$ reads
\begin{equation} \label{zeta_red_eta_deriv0}
    \frac{\pd}{\pd({\eta}^2)} \zeta(s) = \int\limits_0^\infty \frac{k dk}{2\pi}
                                   \sum\limits_{\Pi=\pm 1}\sum\limits_m
                                   \frac{-2s}{(\omega_{k,\Pi,m}^2)^{s+1/2}} \frac{\pd\omega_{k,\Pi,m}}{\pd({\eta}^2)}.
\end{equation}
Let us now find the derivatives $\pd\omega_{k,\Pi,m} / \pd({\eta}^2)$ near the
Maxwell solutions (i.e., those in the $\eta = 0$ case, see
Eq.~\eqref{omega_Maxwell}) for $m = 1,2,3,...$ explicitly showing the
dependencies of the functions on $\eta$ in the following expression:
\begin{eqnarray}
    0  = \frac{d^2}{d{\eta}^2} g_\Pi(\omega_n^2({\eta}^2), \eta)\vert_{\eta = 0} &=&
       \left[ \frac{\pd^2g_\Pi(\omega_n^2({\eta}^2),\eta)}{\pd{\eta}^2} +
              \frac{\pd^2g_\Pi(\omega_n^2({\eta}^2),\eta)}{\pd\omega_n^2}
                                \left(\frac{\pd\omega_n({\eta}^2)}{\pd{\eta}}\right)^2\right.  \nonumber\\
        &+& \left. 2\frac{\pd^2g_\Pi(\omega_n^2({\eta}^2),\eta)}{\pd\omega_n\pd\eta}
                                \frac{\pd\omega_n({\eta}^2)}{\pd{\eta}} +
              \frac{\pd g_\Pi(\omega_n^2({\eta}^2),\eta)}{\pd\omega_n} \frac{\pd^2\omega_n({\eta}^2)}{\pd{\eta}^2}
       \right]_{\eta = 0}.
\end{eqnarray}
Since $\omega_n$ is an even function of $\eta$, its first derivative $\pd
\omega_n / \pd\eta$ vanishes at $\eta = 0$, thus the second and the third
terms vanish, and we obtain
\begin{equation}\label{freqShiftFormula}
   \left.\frac{\pd\omega_n(\eta^2)}{\pd(\eta^2)}\right\vert_{\eta = 0}
   = \left.\frac12 \frac{\pd^2\omega_n(\eta^2)}{\pd \eta^2}\right\vert_{\eta = 0}
   = - \left.\frac12\frac{\pd^2 g_\Pi(\omega_n^2,\eta) / \pd\eta^2}
                         {\pd g_\Pi(\omega_n^2,\eta) / \pd\omega_n}
       \right\vert_{\eta = 0}.
\end{equation}
Now, to find the second $\eta$-derivative, let us treat $g_\Pi$ as a function
of $K_\pm$ rather than a function of $\omega$ and $\eta$. Then, according to
\eqref{vac_spec},
\begin{eqnarray}
  \eta &=& \frac{K_- - K_+}{2}, \quad \omega = \sqrt{K_+ K_-}, \quad g_\Pi(\omega^2,\eta) \equiv \mathcal{G}_\Pi(K_+,K_-),\\
  \frac{\pd}{\pd\eta} &=& \frac{\pd K_+}{\pd \eta} \frac{\pd}{\pd K_+} + \frac{\pd K_-}{\pd \eta} \frac{\pd}{\pd K_-}
                      = -\frac{2 K_+}{K_+ + K_-} \frac{\pd}{\pd K_+} +  \frac{2 K_-}{K_+ + K_-} \frac{\pd}{\pd K_-},\\
  \left.\frac{\pd^2 g_\Pi}{\pd\eta^2}\right\vert_{\eta=0} &=&
  \left.\left[ \left(\frac{\pd}{\pd K_+} - \frac{\pd}{\pd K_-}\right)^2
       + \frac{1}{K_+} \left(\frac{\pd}{\pd K_+} + \frac{\pd}{\pd K_-}\right)\right]\mathcal{G}_\Pi(K_+,K_-)
       \right\vert_{K_+ = K_- = \omega} \nonumber\\
  &=& \frac{d^2\mathcal{G}_\Pi(\omega,\omega)}{d\omega^2} + \frac{1}{\omega}\frac{d\mathcal{G}_\Pi(\omega,\omega)}{d\omega}
  - 4 \left.\frac{\pd^2\mathcal{G}_\Pi}{\pd K_+ \pd K_-}\right\vert_{K_+ = K_- = \omega},\\
  \left.\frac{\pd g_\Pi}{\pd \omega}\right\vert_{\eta = 0} &=& \frac{d\mathcal{G}_\Pi(\omega,\omega)}{d \omega}.
\end{eqnarray}
Taken at the zeros $\omega_{\text{M}}$ of $g_\Pi$ in the $\eta = 0$ case (see
Eq.~\eqref{omega_Maxwell}), these derivatives read
\begin{eqnarray}
    \left.\frac{\pd^2 g_\Pi}{\pd\eta^2}\right\vert_{\eta = 0} &=&
    \frac{(-1)^m 2a}{\omega_{\text{M}}} \left(1 - ( 1 - 2 (-1)^m \Pi) \frac{k^2}{\varkappa_{\text{M}}^2} \right),\\
    \left.\frac{\pd g_\Pi}{\pd \omega}\right\vert_{\eta = 0} &=& (-1)^m 2a,\\
    \omega_{\text{M}} &=& \sqrt{k^2 + \varkappa_{\text{M}}^2}, \quad \varkappa_{\text{M}} = \frac{\pi m}{2a},
    \quad m = 1,2,3,...
\end{eqnarray}
Substituted into \eqref{freqShiftFormula}, this results in the following
expression for the frequency shift:
\begin{eqnarray}
  \left.\frac{\pd\omega_{m,k,\Pi}}{\pd(\eta^2)}\right\vert_{\eta = 0}
  &=& -\frac{1}{2\omega_{\text{M}}} \left(1 - ( 1 - 2 (-1)^m \Pi) \frac{k^2}{\varkappa_{\text{M}}^2} \right),\\
  \left.\sum\limits_{\Pi = \pm1}\frac{\pd\omega_{m,k,\Pi}}{\pd(\eta^2)}\right\vert_{\eta = 0}
  &=& -\frac{1}{\omega_{\text{M}}} \left(1 - \frac{k^2}{\varkappa_{\text{M}}^2} \right) =
      -\frac{1}{\omega_{\text{M}}} \left(1 - \frac{4a^2k^2}{\pi^2m^2} \right). \label{omega_shift}
\end{eqnarray}
Remembering that in the Maxwell case, the frequencies $\omega_{\text{M}}$ do
not depend on the parity $\Pi$, we substitute this result into
\eqref{zeta_red_eta_deriv0} and, after the summation over $\Pi = \pm1$,
obtain:
\begin{eqnarray}
  \left.\frac{\pd}{\pd({\eta}^2)} \zeta(s)\right\vert_{\eta = 0} &=&
                                   \int\limits_0^\infty \frac{k dk}{2\pi}
                                   \sum\limits_{m=1}^\infty
                                   \frac{2s}{\left(k^2 + (\pi m / 2a)^2\right)^{s+1}} \,
                                   \left(1 - \frac{4a^2k^2}{\pi^2m^2} \right) = \frac{s}{\pi}
                                   \left(I(s+1,0) -\frac{4a^2}{\pi^2}I(s+1,2)\right),\\
  \left.\zeta(s)\right\vert_{\eta = 0} &=&
                                   \int\limits_0^\infty \frac{k dk}{2\pi}
                                   \sum\limits_{m=1}^\infty
                                   \frac{2}{\left(k^2 + (\pi m / 2a)^2\right)^s} = \frac{1}{\pi} I(s,0),\\
  I(s,\alpha) &=&                  \int\limits_0^\infty k dk \sum\limits_{m=1}^{\infty}\frac{1}{\left(k^2 + (\pi m / 2a)^2\right)^s}
                                   \left(\frac{k}{m}\right)^\alpha.
\end{eqnarray}
These integrals can be calculated using the integral representation of the
Euler gamma function, the Gauss integral identity, and the series
representation of the Riemann zeta function $\zeta_{\text{R}}(s)$
\cite{RiemannZeta},
\begin{eqnarray}
   \frac{1}{\left(k^2 + (\pi m / 2a)^2\right)^s} &=&
   \frac{1}{\Gamma(s)}\int\limits_0^\infty t^{s-1} e^{-\left(k^2 + (\pi m / 2a)^2\right)t} dt, \quad {\RRe s > 0},
                                                                        \label{identity:EulerGamma}\\
   \int\limits_{0}^\infty k^{\alpha+1} e^{-k^2t} dk &=& \frac{\Gamma(\frac{\alpha}{2}+1)}{2 t^{\frac{\alpha}{2} + 1}},
                                                         \quad \RRe\alpha > -2, \, t > 0, \label{identity:GaussIntegral}\\
   \sum\limits_{m=1}^\infty \frac{1}{m^{2s - 2}} &=& \zeta_{\text{R}}(2s - 2), \quad \RRe{s} > 3/2.    \label{identity:RiemannZeta}
\end{eqnarray}
Let us evaluate the integral $I(s,\alpha)$ using these three identities
referred to above the equality signs where they are used:
\begin{eqnarray}
   I(s,\alpha) &\overset{\eqref{identity:EulerGamma}}{=}&
     \frac{1}{\Gamma(s)}\sum\limits_{m=1}^\infty \int\limits_0^\infty k dk  \left(\frac{k}{m}\right)^\alpha
     \int\limits_0^\infty t^{s-1} e^{-t(k^2 + (\pi m / 2a)^2)} dt
     =\frac{1}{\Gamma(s)}\sum\limits_{m=1}^\infty \frac{1}{m^\alpha}
     \int\limits_0^\infty t^{s-1} e^{-t(\pi m /2a)^2}dt
     \int\limits_0^\infty k^{\alpha+1} e^{-k^2 t} dk \nonumber\\
     &\overset{\eqref{identity:GaussIntegral}}{=}&
     \frac{\Gamma(\frac{\alpha}{2}+1)}{2\Gamma(s)}\sum\limits_{m=1}^\infty \frac{1}{m^\alpha}
     \int\limits_0^\infty t^{s - \frac{\alpha}{2} - 2} e^{-t(\pi m / 2a)^2} dt
     \overset{\eqref{identity:EulerGamma}}{=}
     \frac{\Gamma(\frac{\alpha}{2}+1)\Gamma(s-\frac{\alpha}{2}-1)}{2\Gamma(s)}
     \sum\limits_{m=1}^\infty \frac{1}{m^\alpha}\left(\frac{2a}{\pi m}\right)^{2s-\alpha - 2}\nonumber\\
     &\overset{\eqref{identity:RiemannZeta}}{=}&
     \frac{\Gamma(\frac{\alpha}{2}+1)\Gamma(s-\frac{\alpha}{2}-1)}{2\Gamma(s)}
     \left(\frac{2a}{\pi}\right)^{2s - \alpha - 2} \zeta_{\text{R}}(2s - 2).
\end{eqnarray}
These transformations are valid (i.e., all integrals are convergent) when
$\RRe \alpha > -2$, $\RRe s > \max(\frac{3}{2}, 1 + \RRe\frac{\alpha}{2})$,
i.e., for sufficiently large $\RRe s$ for any fixed $\alpha$, such that $\RRe
\alpha > -2$.

Then, for $\RRe s > \frac{3}{2}$, both the zeta function $\zeta(s)$ and its
$\eta^2$-derivative at $\eta = 0$ are convergent,
\begin{eqnarray}
  \left.\zeta(s)\right\vert_{\eta = 0} &=& \frac{1}{2\pi(s-1)}\left(\frac{2a}{\pi}\right)^{2s-2} \zeta_{\text{R}}(2s-2), \\
  \left.\frac{\pd}{\pd(\eta^2)}\zeta(s)\right\vert_{\eta = 0} &=& \frac{s-2}{2\pi(s-1)} \left(\frac{2a}{\pi}\right)^{2s}
  \zeta_{\text{R}}(2s),\\
  \zeta(s) &=& \frac{1}{2\pi(s-1)} \left(\frac{D}{\pi}\right)^{2s-2} \left(\zeta_{\mathrm{R}}(2s-2) + (s-2)\left(\frac{\eta
      D}{\pi}\right)^2 \zeta_{\mathrm{R}}(2s) + \smallO(\eta^3) \right),  \label{zeta_correction}
\end{eqnarray}
where $D = 2a$ is the distance between the plates, as it was mentioned before.

One should also analyze the contribution of the ``quasi-zero'' modes into the
Casimir effect. These modes formally correspond to $m = 0$ and can be left
aside in the $\eta = 0$ case (see comments after Eq.~\eqref{omega_Maxwell}).
It is easily seen from \eqref{omega_shift} that this formula does not not have
its asymptotical meaning in the $m = 0$ case, since $\varkappa_{\text{M}} =
0$. Proper asymptotical solution of the equation \eqref{spec_cond} near
$\omega = k$ (i.e., for $m = 0$) gives the following expressions for the
energies $\omega_{k,\Pi,0}$ of the ``quasi-zero'' modes with parities $\Pi =
\pm1$:
\begin{align}
  \omega_{k,+1,0} &= k \left(1 - \frac23 \eta^2 a^2 + \bigO(\eta^3)\right), &\quad k \gg \eta,\\
  \omega_{k,-1,0} &= k \pm \eta - \frac{\eta^2}{2k} + \bigO(\eta^3),        &\quad k \gg \eta.
\end{align}
The odd ($\Pi = -1$) solutions do not contribute to the Casimir force within
the second order in $\eta$ since they do not depend on $a$. The even ($\Pi =
+1$) solutions do depend on it, and the corresponding $\eta$-dependent
correction to the zeta function $\zeta(s)$, taken for $k \gg \eta$ solutions,
reads
\begin{equation}
    \eta^2\int\limits_{\Lambda\eta}^\infty \frac{k dk}{2\pi} \frac{-2s}{k^{2s+1}}\left(-\frac23 ka^2\right)
    = \frac{2s \eta^2a^2}{3\pi} \int\limits_{\Lambda\eta}^\infty \frac{d k}{k^{2s -1}} =
    \frac{s}{3\pi(s-1)}\,\frac{\eta^2a^2}{(\Lambda\eta)^{2s - 2}},
    \quad \Lambda \gg 1,
\end{equation}
which has the order $\bigO(\eta^5)$ when $s \to -1/2$. The corrections to the
zeta function due to the domain $k \in [0,\bigO(\eta)]$ have the order
$o(\eta^2)$ due to the smallness of this domain. Thus the ``quasi-zero'' modes
do not contribute to the Casimir force, within the second-order accuracy in
$\eta$. The strict demonstration of this fact will be presented directly in
the next section, where the exact integral expression for the Casimir energy
will be derived. Now we limit ourselves to a qualitative discussion of this
issue.

Using expression \eqref{zeta_correction}, we finally obtain the correction to
the Casimir energy and force,
\begin{eqnarray}
    \frac{E_{\text{vac}}}{L^2} &=& \frac{\zeta(-1/2)}{2}
    = -\frac{\pi^2}{6D^3} \left(\zeta_{\text{R}}(-3) - \frac{5(\eta D)^2}{2\pi^2}\zeta_{\text{R}}(-1) + \smallO((\eta D)^3)
    \right)\nonumber\\
    &=& -\frac{\pi^2}{720 D^3} \left( 1 + \frac{25 (\eta D)^2}{\pi^2} + \smallO((\eta D)^3)\right), \label{E_casimir_zeta}\\
    f_{\text{Casimir}} &=& -\frac{\pd}{\pd D}\frac{E_{\text{vac}}}{L^2} =
    -\frac{\pi^2}{240 D^4} \left( 1 + \frac{25 (\eta D)^2}{3\pi^2} + \smallO((\eta D)^3)\right). \label{F_Casimir_zeta}
\end{eqnarray}
This correction is relevant when the dimensionless parameter $|\eta| D \ll 1$,
i.e., at relatively short distances. It should be stressed that the zeta
function regularization automatically gave us the renormalized result for the
Casimir energy, within the leading order in $\eta$.
\section{The Casimir energy: series summation via the residue theorem}\label{sec:ResidueTheorem}
In this section, we will make a renormalization of the vacuum energy series
\eqref{E_vac} using the direct summation of the one-particle eigenstate
energies implicitly defined as the zeros of the functions $g_\Pi(\omega^2)$
(see Eq.~\eqref{vac_spec}). The explicit expression for the sum over the
energy eigenvalues $\omega_{k,\Pi,m}$ can be derived using the residue theorem
\cite{ComplexAnalysis}.

First, as it was mentioned in Sec.~\ref{sec:SpectrumNotes}, the properties of
the spectrum do not depend on the sign of $\eta$, so we assume that $\eta \ge
0$. Then, it was also shown that the energy eigenvalues are either real or
pure imaginary, and all of them correspond to $K_+ \in [-\eta, \infty)$. The
real eigenvalues correspond to $K_+ \in [0,\infty)$, so the real part of the
vacuum energy can be presented in the following form:
\begin{eqnarray}\label{Re_Evac}
  \frac{1}{L^2}\RRe E_{\text{vac}} &=& \frac12\int \frac{k dk}{2\pi} D \sum\limits_{\Pi = \pm1} S_\Pi(D), \\
  S_\Pi(D) &=& \frac{1}{D}{\sum\limits_m}' \omega_{k, m, \Pi},
\end{eqnarray}
where $\sum_m'$ means a summation over those $m$, for which $\omega_{k,m,\Pi}
\in (0,\infty)$, i.e., the corresponding $K_+ \in (0,\infty)$. Let us apply a
regularization and redefine $S_\Pi$ as
\begin{equation}\label{S_Pi}
    S_\Pi(D) = \frac{1}{D} {\sum\limits_m}' \omega_{k, m, \Pi} e^{-\omega_{k,m,\Pi}/\sqrt{k \Lambda}},
\end{equation}
where $\Lambda \to +\infty$ is the cutoff parameter. This sum is convergent
when $\Lambda$ is finite and can be represented as a complex plane contour
integral. Recall the functions introduced in section \ref{sec:SpectrumNotes}
instead of $g_\Pi(\omega^2)$:
\begin{equation}\label{g_tilde2}
        \tilde{g}_\Pi(\omega^2) = \frac{g_\Pi(\omega^2)}{\varphi_\Pi(\varkappa_+ a) \varphi_\Pi(\varkappa_- a)} =
                \tan^\Pi{\varkappa_+ a} \ \sin\theta_+ + \tan^\Pi{\varkappa_- a} \ \sin\theta_-.
\end{equation}
\begin{figure}[tbh]
   \begin{center}
        \includegraphics[width=8.6cm]{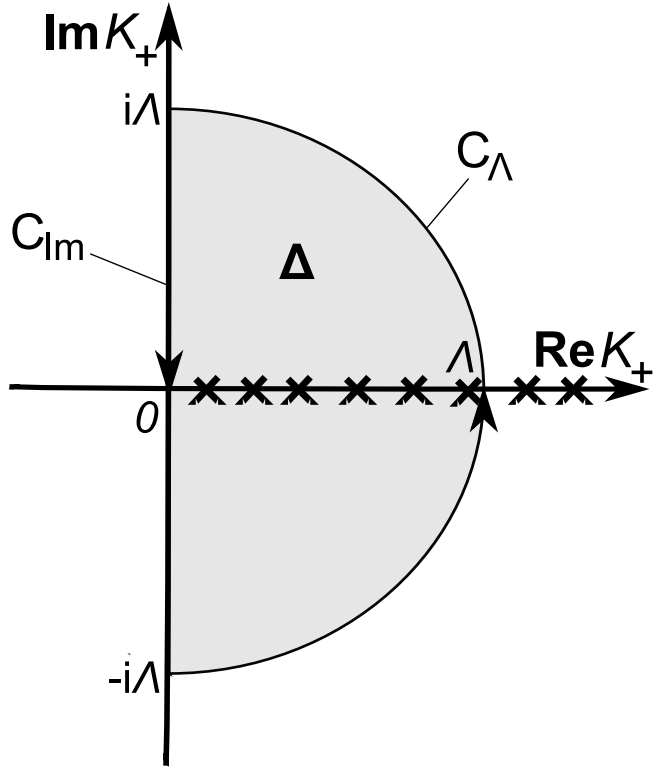}
        \caption{To the transformation of the sum into a complex plane integral.
                 The crosses indicate the zeros of the function $\tilde{g}_\Pi$,
                 which correspond to the real energy eigenvalues.}
        \label{fig:cplxContour}
   \end{center}
\end{figure}
These functions can be treated as functions of $K_+ \in \Complex$ (for fixed
$k \ge 0$, as we will further assume). Although $\varkappa_\pm = \sqrt{K_\pm^2
- k^2}$ have two branches (namely, they are defined with an arbitrary sign),
the expressions $\tan^\Pi{\varkappa_\pm a} \sin\theta_\pm = \frac{1}{K_\pm}
\varkappa_\pm \tan^\Pi\varkappa_\pm a$ have only one branch, moreover,
functions $\tilde{g}_\Pi$ have only pole singularities, i.e., they are
\emph{meromorphic} functions \cite{ComplexAnalysis}. Note that the original
functions $g_\Pi$ were \emph{multivalued} complex functions of $K_+$, so they
were incompatible with the residue theorem.

Since functions $\varphi_\Pi(\varkappa_\pm)$ are bounded in any finite circle
$\{|K_+| < M\}$, $M > 0$, the set of zeros of $\tilde{g}_\Pi$ lies within the
set of zeros of $g_\Pi$. It implies, in particular, that the zeros of the
functions $\tilde{g}_\Pi$ also correspond to real values of $K_+$, as it was
for the zeros of $g_\Pi$. Moreover, it can be explicitly shown that the
difference between the two abovementioned sets contains only those points for
which either $K_+ \in \{0, \pm k\}$, or $K_- \in \{ 0, \pm k \}$, or
$\varphi_\Pi(\varkappa_+ a) = \varphi_\Pi(\varkappa_- a) = 0$. The first two
cases correspond to the energy eigenvalues that do not depend on $a$ and
therefore can be excluded from the Casimir energy series. On the other hand,
the solutions of the equation $\varphi_\Pi(\varkappa_+ a) =
\varphi_\Pi(\varkappa_- a) = 0$ \emph{do not exist for almost all $k$} (all
but no more than a countable set), so the summation over these additional
energy eigenvalues gives a vanishing contribution to the vacuum energy, after
the integration over $k$. Thus it is convenient to think of $S_\Pi$ as of
\emph{the sum over all zeros of the function $\tilde{g}_\Pi(\omega^2)$}
corresponding to $K_+ \in (0, \infty)$, regularized in the way shown in
\eqref{S_Pi}.

Now we can apply the residue theorem to the domain $\Delta = \{ \RRe K_+ > 0,
|K_+| < \Lambda \}$ (see Fig.\ref{fig:cplxContour}). Denote the contour
enclosing this domain as $C \equiv C_{\text{Im}} + C_\Lambda$, where
$C_{\text{Im}}$ corresponds to the segment of the imaginary axis, and
$C_\Lambda$ is the semicircle with the radius $\Lambda$, as shown in
Fig.\ref{fig:cplxContour}. The direction of this contour is chosen
``mathematically-positive'', i.e., counterclockwise. Then the residue theorem
implies that
\begin{eqnarray}\label{residueTheorem}
    \oint\limits_C \frac{dK_+}{2\pi i} \, \omega(K_+) e^{-\omega(K_+)/\sqrt{k\Lambda}}
    \, \frac{\pd \tilde{g}_\Pi / \pd K_+}{\tilde{g}_\Pi} &=&
    \sum\limits_{K_+^{(n)}}\omega(K_+^{(n)}) e^{-\omega(K_+^{(n)})/\sqrt{k\Lambda}}
     \nonumber\\
    &+&\sum\limits_{K_+^{(p)}} \omega(K_+^{(p)})e^{-\omega(K_+^{(p)})/\sqrt{k\Lambda}}\,
    \Res{\frac{\pd \tilde{g}_\Pi / \pd K_+}{\tilde{g}_\Pi}}{\,K_+ = K_+^{(p)}},
\end{eqnarray}
where $\omega(K_+) = \sqrt{K_+ K_-} = \sqrt{K_+ (K_+ + 2\eta)}$ is the
analytical function in $\Delta$. $K_+^{(n)}$ correspond to the zeros of
$\tilde{g}_\Pi$ lying within $\Delta$, while $K_+^{(p)}$ correspond to the
poles of the numerator, i.e., $\pd\tilde{g}_\Pi / \pd K_+$. The first sum in
the right side of \eqref{residueTheorem} can be replaced by $S_\Pi D$ when
$\Lambda \to \infty$, since we have shown in the previous paragraph that we
can make a summation over the zeros of the function $\tilde{g}_\Pi$ instead of
$g_\Pi$. Let us substitute the functions $\tilde{g}_\Pi$ and their derivatives
\begin{equation}
   \frac{1}{a} \frac{\pd \tilde{g}_\Pi}{\pd K_+} = \sum\limits_{\lambda=\pm1}
   \left(\Pi (1 + \tan^{2\Pi}\varkappa_\lambda a)
   + \frac{\tan^\Pi\varkappa_\lambda a}{\varkappa_\lambda a} \cos^2\theta_\lambda\right)
\end{equation}
into \eqref{residueTheorem}:
\begin{eqnarray}
   S_\Pi &=& \frac{1}{2a} \oint\limits_C \frac{dK_+}{2\pi i} \, \omega(K_+) e^{-\omega(K_+)/\sqrt{k\Lambda}}\,
   \frac{\pd \tilde{g}_\Pi / \pd K_+}{\tilde{g}_\Pi}
   - \frac{1}{2} \sum\limits_{K_+^{(p)}} \omega(K_+^{(p)}) e^{-\omega(K_+^{(p)})/\sqrt{k\Lambda}} \nonumber\\
   &\times&\Res{\frac{\sum\limits_{\lambda=\pm1}
   \left(\Pi (1 + \tan^{2\Pi}\varkappa_\lambda a)
   + \frac{\tan^\Pi\varkappa_\lambda a}{\varkappa_\lambda a} \cos^2\theta_\lambda\right)}
   {\sin\theta_+ \tan^\Pi\varkappa_+ a + \sin\theta_- \tan^\Pi\varkappa_- a}}{K_+ = K_+^{(p)}}.  \label{S_Pi_1}
\end{eqnarray}
One can see that the summation is performed over those values of $K_+$  for which either $\tan^\Pi\varkappa_+ a
\to \infty$ or $\tan^\Pi\varkappa_- a \to \infty$, and the residue is equal to the residue of the function
$\Psi_\Pi(K_+) = \Pi\left(\frac{\tan^\Pi\varkappa_+ a}{\sin\theta_+} + \frac{\tan^\Pi\varkappa_-
a}{\sin\theta_-}\right)$, which does not have any singular points other than $K_+^{(p)}$. Then the sum over the
poles $K_+^{(p)}$ in the right side of \eqref{S_Pi_1} can be also presented as an integral over the contour $C$,
with an integrand containing $\Psi_\Pi(K_+)$. Then the two integrals, taken together, give
\begin{eqnarray}
  S_\Pi &=&
  \frac{1}{2a} \oint\limits_C \frac{dK_+}{2\pi i} \, \omega(K_+) e^{-\omega(K_+)/\sqrt{k\Lambda}}\,
  \left\{ \frac{\pd \tilde{g}_\Pi / \pd K_+}{\tilde{g}_\Pi}
       - a \Psi_\Pi(K_+)
  \right\} \nonumber\\
  &=&
  \frac{\Pi}{2} \oint\limits_C \frac{d K_+}{2\pi i} \omega(K_+)
  e^{-\omega(K_+)/\sqrt{k\Lambda}}\frac{\Xi_\Pi}{\tilde{g}_\Pi},\\
  \Xi_\Pi &=&\sum\limits_{\lambda=\pm1}\left(1 - \tan^\Pi\varkappa_\lambda a \tan^\Pi{\varkappa_{-\lambda}a}
                     \frac{\sin\theta_\lambda}{\sin\theta_{-\lambda}}
               + \frac{\Pi\tan^\Pi\varkappa_\lambda a\cos^2\theta_\lambda}{\varkappa_\lambda a}\right)
\end{eqnarray}
Now we need to renormalize the integrals we have obtained, i.e., subtract
their $\Lambda \to \infty$ divergencies, which can be extracted analyzing the
case $a \to \infty$, i.e., infinitely distant plates. Two types of divergent
counterterms should be subtracted, namely, in terms of $S_\Pi$,
\label{page:physRenorm}
\begin{equation}\label{S_Pi_div}
    S_\Pi(D,\Lambda) \to S_\Pi^{\text{ren}}(D,\Lambda)
    = S_\Pi(D,\Lambda) - \frac{C^{(1)}_\Pi(\Lambda)}{D} - C^{(2)}_\Pi(\Lambda),
\end{equation}
where $C^{(1,2)}_\Pi$ are the functions of $k$, $\eta$, and the regulator $\Lambda$, but they should not depend
on the distance $D$ between the plates. The term with $C^{(1)}_\Pi$ gives a constant contribution to the vacuum
energy, so it is non-physical. The other term, which contains $C^{(2)}_\Pi$, gives a constant contribution to
the ``average vacuum energy density'', that is, the vacuum energy divided by $L^2D$. It is obvious that, for
large $D$, this ``vacuum energy density'' is equal to the force acting on a plate in a semispace. Thus
subtracting it means the subtraction of the force acting upon the open side of a plate, which is physically
reasonable.

First, let us analyze the asymptotic behavior of the integral over the semicircle $C_\Lambda$  for $\Lambda \to
\infty$. When $K_+ = \Lambda e^{i\vartheta}$, $|\vartheta| < \pi/2 - \epsilon$, $\epsilon > 0$, the integrand is
exponentially small due to the factor $e^{-\omega / \sqrt{k\Lambda}}$ ($\omega(K_+)$ is a one-valued function in
$\Delta$, with $\RRe\omega > 0$). On the other hand, when $\vartheta$ is near $\pm\pi/2$ so that $\RRe\omega
\not\gg \sqrt{k\Lambda}$, we can use another set of approximations:
\begin{eqnarray}
  \omega, K_\pm, \varkappa_\pm  &=& \bigO(\Lambda),                \label{approx_C_Lambda_first}\\
  \sin\theta_\pm &=& 1 + \bigO(1/\Lambda^2), \\
  \cos\theta_\pm &=& \bigO(1/\Lambda), \\
  \tan^\Pi\varkappa_\pm a &=& i \Pi \sgn \vartheta + \bigO(e^{-2\Lambda a}), \label{approx_C_Lambda_last}
\end{eqnarray}
where $K_- = K_+ + 2\eta$. Then the integrand
\begin{equation}\label{integrand1}
    \omega(K_+) e^{-\omega(K_+) /\sqrt{k\Lambda}}\, \frac{\Xi_\Pi}{\tilde{g}_\Pi} =
    \omega(K_+) (-4i\Pi\sgn\vartheta + \bigO(1/\Lambda^2)) e^{-\omega(K_+) /\sqrt{k\Lambda}},
\end{equation}
moreover, the terms entering this expression that depend on $a$ (in other words, \emph{change} of this
expression if it is treated as a function of $a$) have the order $\bigO(1/\Lambda^2)$. Since $\omega(K_+) =
\bigO(\Lambda)$ and the integration effectively includes the pieces of $C_\Lambda$ of length
$\bigO(\sqrt{k\Lambda})$, where the exponential factor is not infinitely small, the integral over $C_\Lambda$
has the order $\bigO(\sqrt{k\Lambda^3})$ while \emph{its parts that depend on $a$} have the order
$\bigO(\sqrt{k/\Lambda})$ and \emph{vanish in the $\Lambda \to \infty$ limit}. The latter fact holds true even
for $a\to \infty$. Thus the integral over $C_\Lambda$ has the form of the counterterm containing $C_\Pi^{(2)}$,
except for the terms that vanish in the $\Lambda \to \infty$ limit. Hence, \emph{this integral is fully
cancelled when renormalized}.

Second, let us consider the $a \to \infty$ asymptotic of the integral over
$C_{\text{Im}}$ to extract the counterterms. In this limit,
\begin{equation}
    \tan^\Pi \varkappa_\pm a \to i\Pi \sigma, \quad \sigma \equiv \sgn\IIm K_+,
\end{equation}
where the difference between the left and the right sides is exponentially
small in $a$, i.e., smaller than any negative power of $a$. Then the integrand
takes the following asymptotical form:
\begin{equation}\label{integrand_div}
    \frac{\Xi_\Pi}{\tilde{g}_\Pi} = -i\Pi\sigma \left(\frac{1}{\sin\theta_+} + \frac{1}{\sin\theta_-}\right)
                + \frac{\Pi}{a(\sin\theta_+ + \sin\theta_-)}
                \left( \frac{\cos^2\theta_+}{\varkappa_+} + \frac{\cos^2\theta_-}{\varkappa_-} \right) +
                \text{ $\exp$-small terms in $a$}.
\end{equation}
Here, we do not need to strictly specify what the exponential smallness in $a$
means (e.g., write expressions of the form $\bigO(e^{-2 a |\IIm
\varkappa_\pm|})$), since our asymptotical treatment is only aimed at
extracting the relevant integral representation of the counterterm, which
should be compatible with \eqref{S_Pi_div}. Namely, we see that the first two
terms in the above expression contribute to $C^{(2)}_\Pi$ and $C^{(1)}_\Pi$,
respectively. Hence, we subtract these two terms, multiplied by $\omega
e^{-\omega / \sqrt{k\Lambda}}$, from the integrand $\omega e^{-\omega /
\sqrt{k\Lambda}} \Xi_\Pi / \tilde{g}_\Pi$ \emph{for finite $a$}. Since the
integrand over $C_\Lambda$ vanishes after renormalization and setting $\Lambda
\to \infty$, then, after this subtraction, we obtain the \emph{integral
representation of the renormalized sum over the frequencies}:
\begin{eqnarray}
  S_\Pi^{\text{ren}} &=& \frac{\Pi}{2} \int\limits_{C_{\text{Im}}} \frac{d K_+}{2\pi i}
  \omega(K_+) e^{-\omega(K_+)/\sqrt{k\Lambda}}
  \frac{\Xi_\Pi^{\text{ren}}}{\sin\theta_+ \tan^\Pi\varkappa_+ a + \sin\theta_- \tan^\Pi\varkappa_- a},\\
  \Xi_\Pi^{\text{ren}} &=&  2 -
             \sum\limits_{\lambda = \pm1} \tan^\Pi{\varkappa_\lambda a}
                \left\{
                         \tan^\Pi \varkappa_{-\lambda}a \frac{\sin\theta_\lambda}{\sin\theta_{-\lambda}}
                         - i\Pi\sigma \left(1+\frac{\sin\theta_\lambda}{\sin\theta_{-\lambda}}\right)
                         - \sum\limits_{\lambda' = \pm1}
                               \frac{\Pi \lambda\lambda' \cos^2\theta_{\lambda'} \sin\theta_{-\lambda'}}
                                    {(\sin\theta_+ + \sin \theta_-)\varkappa_{\lambda'} a}
                \right\}.
\end{eqnarray}
One can easily show that the integrand is exponentially small for $|\IIm K_+|
\gg \eta, 1/a$, \emph{even without the regularizing factor
$e^{-\omega/\sqrt{k\Lambda}}$}. Moreover, the integral is convergent near $K_+
= 0$. Now we are able to set the regulator $\Lambda$ to infinity, then the
smooth cutoff factor $e^{-\omega /\sqrt{k\Lambda}}$ disappears. The
integration is performed over the imaginary axis $K_+ \in C_{\text{Im}} = (+i
\infty, -i\infty)$, so it is natural to make a reparametrization making $K_+$
real (namely, $K_+ \to iK_+$, $dK_+/2\pi i \to -dK_+/2\pi$, $K_+ \in \Reals$).
Moreover, looking at the expression \eqref{Re_Evac} for the vacuum energy and
the expression for $S_\Pi^{\text{ren}}$, we see that we can also make a
reparametrization of the integration variables making them dimensionless, such
as $k a$ and $K_+ a$. Making these two types of reparametrizations, we arrive
at the final expression for the Casimir energy:
\begin{eqnarray}
   \frac{1}{L^2}\RRe E^{\text{ren}}_{\text{vac}} &=& -\frac{1}{4\pi a^3}\int\limits_0^\infty\frac{k dk}{2\pi}
                     \sum\limits_{\Pi = \pm1}
                     \int\limits_{-\infty}^\infty d K_+\, \omega(K_+)
                     \frac{\Xi^{\text{ren}}_\Pi}{\tilde{g}_\Pi},    \label{E_vac_ren}\\
  \Xi_\Pi^{\text{ren}} &=&  2 - \sum\limits_{\lambda=\pm1}
                                                   \tanh^\Pi\varkappa_\lambda
                                                   \left\{
                                                      1
                                                      + \frac{\cosh\theta_\lambda}{\cosh\theta_{-\lambda}}
                                                           (1 - \tanh^\Pi\varkappa_{-\lambda})
                                                      + \sum\limits_{\lambda' = \pm1}
                                                           \frac{\lambda\lambda' \sinh^2\theta_{\lambda'}
                                                                                 \cosh\theta_{-\lambda'}}
                                                                {\varkappa_{\lambda'}(\cosh\theta_+ + \cosh \theta_-)}
                                                   \right\},\\
  \tilde{g}_\Pi &=& \cosh\theta_+ \tanh^\Pi \varkappa_+ + \cosh\theta_- \tanh^\Pi\varkappa_-,\\
  K_- &=& K_+ - 2 i \eta a, \\
  \omega(K_+) &=& \sqrt{K_+ K_-},\\
  \varkappa_\pm &=& \sqrt{K_\pm^2 + k^2}, \\
  \sinh\theta_\pm &=& \frac{k}{K_\pm}, \quad  \cosh\theta_\pm = \frac{\varkappa_\pm}{K_\pm},
\end{eqnarray}
where $k, K_\pm, \omega, \varkappa_\pm$ are now dimensionless, and all the
square roots are taken in the algebraic sense, i.e., with a branch cut over
negative real numbers and $\RRe \sqrt{x} \ge 0 \,\,\, \forall x \in \Complex
\backslash (-\infty, 0)$. The integrand exponentially falls down with the
increase of $\varkappa_+ = \sqrt{K_+^2 + k^2}$, so the integral is convergent
at infinity. It is easy to show that it is also convergent near zero.

The integrand is complex-conjugate for the opposite values of $K_+$, then the
result is real, as should be. Moreover, the only imaginary quantity in the
integrand expression is $2 i \eta a$, so, to make the integral real, only even
powers of $\eta$ should be present in the result. This means that the same
integral as we have obtained above, applies to the case $\eta < 0$ (although,
in the above calculations, we assumed that $\eta \ge 0$). The $\eta$ parameter
is present as a dimensionless combination $\eta a$.

In the Maxwell ($\eta = 0$) case, we obtain the following integral:
\begin{equation}\label{E_vac_ren_eta_0}
  \left.\frac{1}{L^2} E_{\text{vac}}^{\text{ren}} \right\vert_{\eta = 0} =
                   - \frac{1}{4\pi^2 a^3} \int\limits_0^\infty k dk
                   \int\limits_{-\infty}^\infty K_+^2 dK_+ \frac{(1 - \tanh\varkappa)^2}{\varkappa\tanh\varkappa},
\end{equation}
where $\varkappa = \sqrt{K_+^2 + k^2}$. Using the polar coordinates
$\varkappa, \xi$, so that $k = \varkappa \cos\xi$, $K_+ = \varkappa\sin\xi$,
$\xi \in [-\pi/2,\pi/2]$, and making the integration over $\xi$, we obtain
\begin{equation}\label{E_vac_ren_eta_0_result}
    \left.\frac{1}{L^2} E_{\text{vac}}^{\text{ren}} \right\vert_{\eta = 0} =
    -\frac{1}{6\pi^2 a^3} \int\limits_0^\infty \varkappa^3 d\varkappa
    \frac{(1-\tanh\varkappa)^2}{\tanh\varkappa} =
    -\frac{2}{3\pi^2 a^3} \frac{\Gamma(4)\zeta_{\text{R}}(4)}{4^4} = -\frac{\pi^2}{5760a^3} = -\frac{\pi^2}{720
    D^3},
\end{equation}
which leads to the classical result
\begin{equation}
    \left.f_{\text{Casimir}}\right\vert_{\eta = 0} =
    -\left.\frac{\pd}{\pd D}\frac{1}{L^2} E_{\text{vac}}^{\text{ren}} \right\vert_{\eta = 0} = -\frac{\pi^2}{240
    D^4}.
\end{equation}

To find the corrections to this force, one should take partial derivatives of
the integrand in the expression \eqref{E_vac_ren} with respect to $\eta$. This
integrand, as seen, is symmetric as a function of $K_+, K_-$. Then one can
easily demonstrate that
\begin{equation}
   \left.\frac{\pd}{\pd\eta}\left(\omega(K_+) \frac{\Xi_\Pi^{\text{ren}}}{\tilde{g}_\Pi}\right)\right\vert_{\eta = 0}
   = -i a \left.\frac{d}{d K_+}\left(\omega(K_+) \frac{\Xi_\Pi^{\text{ren}}}{\tilde{g}_\Pi}\right)\right\vert_{\eta = 0},
\end{equation}
i.e., the partial derivative with respect to $\eta$ is proportional to the total derivative with respect to
$K_+$. The latter one vanishes after the integration over $K_+$. The expression for the second derivative we are
interested in is much more complicated, so we present it in a slightly different form, namely, when $\eta$ is
set to zero after the differentiation, we use the polar coordinates $\varkappa, \xi$ (see above), since they are
separated in the $\eta = 0$ case. This second derivative is real and has the following form:
\begin{eqnarray}
  &&\frac{\pd^2}{\pd(\eta a)^2} \,\left.\omega(K_+) \sum\limits_{\Pi = \pm1}
  \frac{\Xi_\Pi^{\text{ren}}}{\tilde{g}_\Pi}\right\vert_{\eta = 0} =
  \frac{1}{2\varkappa \sinh^2 2\varkappa} \left\{ -24 \varkappa^2 \coth 2\varkappa  + (8 \varkappa - 7) \cosh 4 \varkappa
  +  3 \sinh 4\varkappa + 16\varkappa + 7 \right\} - 8 \varkappa + 4 \nonumber\\
  &+& \frac{2\cos 2\xi}{\varkappa} \left\{ \frac{-2\varkappa\sinh 4\varkappa - \cosh 4\varkappa
  + 8\varkappa^2 + 1}{\sinh^2 2\varkappa \tanh 2\varkappa} + 4\varkappa + 2\right\} \nonumber\\
  &+& \cos 4\xi \, \frac{1 - \coth 2\varkappa}{2\varkappa\sinh^2 2\varkappa}
  \left\{4\varkappa(\varkappa+1)\sinh 4\varkappa +
  (4\varkappa^2 + 4\varkappa + 3)\cosh 4\varkappa + 4\varkappa^2 - 4\varkappa - 3\right\}.
\end{eqnarray}
After integrating over $\xi \in [-\pi/2, \pi/2]$, we obtain
\begin{equation}
   \int\limits_{-\pi/2}^{\pi/2} \cos\xi d\xi \frac{\pd^2}{\pd(\eta a)^2} \,\left.\omega(K_+) \sum\limits_{\Pi = \pm1}
  \frac{\Xi_\Pi^{\text{ren}}}{\tilde{g}_\Pi}\right\vert_{\eta = 0}
  =
  \frac{2}{15\varkappa \sinh^2 2\varkappa} (144 \varkappa \sinh 2\varkappa + 7\cosh 6\varkappa -
  (96 \varkappa^2 + 7)\cosh 2\varkappa) - \frac{56}{15\varkappa}.
\end{equation}
Finally, the second derivative of the vacuum energy with respect to $\eta$
reads
\begin{gather}
  \left.\frac{\pd^2}{\pd(\eta a)^2}\frac{\RRe E_{\text{vac}}^{\text{ren}}}{L^2}\right\vert_{\eta = 0}
  = -\frac{1}{8\pi^2 a^3} \int\limits_0^\infty k dk \int\limits_{-\infty}^{\infty} dK_+
  \frac{\pd^2}{\pd(\eta a)^2} \,\left.\omega(K_+) \sum\limits_{\Pi = \pm1}
  \frac{\Xi_\Pi^{\text{ren}}}{\tilde{g}_\Pi}\right\vert_{\eta = 0}  \nonumber\\
  = -\frac{1}{8\pi^2 a^3} \int\limits_0^\infty \varkappa^2 d\varkappa
  \int\limits_{-\pi/2}^{\pi/2} \cos\xi
   d\xi \frac{\pd^2}{\pd(\eta a)^2} \,\left.\omega(K_+) \sum\limits_{\Pi = \pm1}
   \frac{\Xi_\Pi^{\text{ren}}}{\tilde{g}_\Pi}\right\vert_{\eta = 0} \nonumber\\
  = -\frac{1}{8\pi^2 a^3} \int\limits_0^\infty \varkappa d\varkappa
  \left\{\frac{2}{15\sinh^2 2\varkappa} (144 \varkappa \sinh 2\varkappa + 7\cosh 6\varkappa -
  (96 \varkappa^2 + 7)\cosh 2\varkappa) - \frac{56}{15} \right\} = -\frac{5}{144 a^3}. \label{E_vac_ren_eta_2}
\end{gather}
Note that we have proved in Sec.~\ref{sec:SpectrumNotes} that for $|\eta| <
\pi / 4a$, the imaginary-energy solutions exist only for $k = |\eta|$ and they
do not contribute to the vacuum energy which contains the integral over
momentum $k$. Hence we can write $E_{\text{vac}}^{\text{ren}}$ instead of
$\RRe E_{\text{vac}}^{\text{ren}}$ in the $|\eta| \ll 1/a$ case. In this case,
taking \eqref{E_vac_ren_eta_0_result} and \eqref{E_vac_ren_eta_2} together, we
find the approximate expression for the Casimir energy and the Casimir force
in the Maxwell-Chern-Simons electrodynamics:
\begin{eqnarray}
    \frac{1}{L^2}E_{\text{vac}}^{\text{ren}} &=& -\frac{\pi^2}{720 D^3} \left(1 + \frac{25(\eta D)^2}{\pi^2} +
                                                 \smallO((\eta D)^3)\right),\\
    f_{\text{Casimir}} = -\frac{\pd}{\pd D}\frac{1}{L^2}E_{\text{vac}}^{\text{ren}} &=&
    -\frac{\pi^2}{240 D^4} \left( 1 + \frac{25 (\eta D)^2}{3\pi^2} + \smallO((\eta D)^3)\right), \quad |\eta| D \ll 1,
\end{eqnarray}
which are precisely the same as those obtained using the zeta function
approach \eqref{E_casimir_zeta}, \eqref{F_Casimir_zeta}. However, here we have
strictly accounted for the ``quasi-zero'' modes which were discussed
qualitatively in the end of Sec.~\ref{sec:ZetaReg}. Our expression
\eqref{E_vac_ren} for the real part of the Casimir energy is exact for large
$|\eta| D$, i.e., at large distances.

\section{Discussion and conclusion}\label{sec:Conclusion}
Let us briefly discuss the results of our calculations. Originally, M.~Frank and I.~Turan attempted to solve the
problem we discussed here, using the Green's function method, in \cite{Turan}. But they have used a wrong
identity, namely, $(-\pd^2 + \eta^2) \epsilon^{\mu\nu\alpha\beta} \pd_\alpha A_\beta = 0$, in the MCS
electrodynamics, that has reduced the dispersion relation for the MCS photon to that for a massive photon. The
fact is, in 4 dimensions, the existence of the Chern-Simons term does not make the photons just massive, like it
is in the 3-dimensional Maxwell-Chern-Simons electrodynamics \cite{Milton}, instead, the photon possesses a more
complicated dispersion relation (see Eq.~\eqref{vac_spec}). The result we have obtained differs both in sign and
in magnitude from the one obtained in \cite{Turan}. It should be also mentioned that the calculation of the
Green function within the MCS electrodynamics seems to be much more complicated than it was thought in
\cite{Turan}, so in the present paper we have used the two other methods, namely, the zeta function
regularization and the summation and renormalization of the discrete sum involving the residue theorem.

The zeta function analytical regularization is widely used for the
calculations of the Casimir effect in various physical situations
\cite{MiltonLectures}. It automatically subtracts the vacuum energy of the
infinite space (i.e., without the plates) from the Casimir energy. In our
calculations, we used the ``true'' zeta function regularization, in which the
complex regularization parameter $s$, it depends on, controls the negative
power of the frequencies in the series \eqref{zeta_red}. Sometimes (see, e.g.,
\cite{MiltonLectures}), the space dimensionality $d$ is chosen as the
parameter of the analytical regularization, and, indeed, the expression for
the vacuum energy depends on $d$ through the Riemann zeta function. In our
case, it would not be strict enough, since the transformations we would make
during the calculation of the vacuum energy (analogous to
\eqref{identity:EulerGamma}, \eqref{identity:GaussIntegral},
\eqref{identity:RiemannZeta}) would not converge together for any $d \in
\Complex$. The regularization with the parameter $s$ avoids this problem, as
it was mentioned in Sec.~\ref{sec:ZetaReg}.

The method we have used to find the sum of a discrete series using the complex
plane integral, is a kind of generalization of the Abel-Plana formula which is
also widely  used in Casimir effect calculations \cite{Mostepanenko}. Indeed,
one can generalize it to find the explicit integral expression for the series
over the roots of a transcendental equation, including, e.g., Bessel functions
\cite{Saharyan}. The approach we have used in section \ref{sec:ResidueTheorem}
is another generalization of this type, however, it does not follow the
approach developed in \cite{Saharyan}.

One should hold in mind that the calculations presented in our paper cannot
provide a direct way to make experimental predictions. For the experimental
purposes, we should take into account the finite conductivity of the plates
and the dispersion of their conductivity, as well as some other aspects. Here
we can only say how the presence of a small $\eta$ condensate, which violates
Lorentz invariance, affects the dependency of the Casimir force on the
distance between the plates.

Again, we conclude that the correction is quadratic in $\eta$ and strengthens the attraction between the plates
at relatively large distances of the order $D \lesssim 1 / |\eta|$. The relative magnitude of this additional
force compared to the Maxwell Casimir force increases quadratically with distance $D$, so large-separation
Casimir effect measurements could give tighter constraints on $\eta$.

Recent observations \cite{Observ} showed the $1-2$\% agreement between the
experiment and the theory of the Casimir effect based on the conventional
Standard Model, at distances of about $R \sim 100...500$ nanometer. Thus we
can conclude that the leading $\eta$-correction to the Casimir force is less
or about $1-2$\% of its value for $\eta = 0$. This leads to the following
constraint:
\begin{equation}
    |\eta^0| \lesssim \sqrt{1\%} \cdot (500 \text{nm})^{-1} \approx 5\cdot10^{-2}\text{eV}.
\end{equation}
Some experiments measure the Casimir effect at distances of several
micrometers, with about a 10\% accuracy \cite{LargeDistanceObserv}, and this
can at least make the above constraint $3-4$ times tighter.

Though this constraint is very loose compared to those astrophysical and some other observations place on
$\eta$, it demonstrates a property of a quantum vacuum within the Extended Standard Model. Moreover, this is the
first constraint placed on the Chern-Simons Lorentz-violating term based on the \emph{correct} calculation of
the Casimir effect in (3+1) dimensions.

\section*{Acknowledgements}
The authors of the paper are grateful to A.~V.~Borisov and A.~E.~Lobanov for useful discussions and remarks.

\end{document}